\documentclass{article}
\pdfoutput=1

\usepackage[left=1in,top=1in,right=1in,bottom=1in,nohead,footskip=15pt]{geometry}

\usepackage[pdftex, colorlinks=true, urlcolor=MyDarkBlue, citecolor=MyDarkRed, linkcolor=MyDarkGreen]{hyperref}
\usepackage[utf8]{inputenc}
\usepackage{amsmath,amsthm,epsf,amscd,amssymb,verbatim}
\usepackage[square,sort,comma,numbers]{natbib}
\usepackage{graphicx}
\usepackage{mathrsfs}
\usepackage{color}
\usepackage{titlesec}
\usepackage{enumerate}
\usepackage[shortlabels]{enumitem}
\usepackage{datetime}
\usepackage[normalem]{ulem}
\usepackage[font=footnotesize]{caption}
\usepackage{abstract}
\usepackage{lipsum}

\usepackage{chemarr}
\usepackage{extarrows}
\usepackage[all]{xy}

\usepackage{subfig}
\usepackage{multirow}
\usepackage{array}
\usepackage{enumitem}

\usepackage{longtable}
\usepackage{wrapfig}

\usepackage{color}
\definecolor{MyDarkRed}{rgb}{0.5,0,0.1}
\definecolor{MyDarkBlue}{rgb}{0.1,0.1,0.5}
\definecolor{MyDarkGreen}{rgb}{0.1,0.5,0.1}
\definecolor{MyRed}{rgb}{1.0,0,0}
\definecolor{MyBlue}{rgb}{0,0,1.0}
\definecolor{MyGreen}{rgb}{0,0.8,0}
\definecolor{lightgray}{rgb}{0.96,0.96,0.96}
\definecolor{darkgray}{rgb}{0.4,0.4,0.4}
\definecolor{gray}{rgb}{0.5,0.5,0.5}

\newtheorem{lemma}{Lemma}
\newtheorem{cor}{Corollary}
\newtheorem{prop}{Proposition}

\theoremstyle{definition}
\newtheorem{definition}{Definition}
\theoremstyle{remark}

\newtheorem{mynote}{Note}

\newcommand{\wipe}[1]{}

\newcommand{\vs}{\mathsf{v}}
\newcommand{\os}{\mathsf{o}}
\newcommand{\us}{\mathsf{u}}

\definecolor{MyRed}{rgb}{1.0,0,0}
\definecolor{MyBlue}{rgb}{0,0,1.0}
\definecolor{MyPurple}{rgb}{0.5,0,0.5}

\newcounter{pathnum}
\newcounter{condnum}[pathnum]

\newcounter{algorithm}
\newenvironment{algorithm}[1]
{ 
  \vspace{1em}
  \refstepcounter{algorithm}
  \begin{tabular}{r|l} 
  \multicolumn{2}{c}{\textbf{Algorithm~\arabic{algorithm}: {#1}}} \\
  \hline
}
{ 
  \end{tabular}
}

\newcommand{\algcomment}[1]{{\small\color{gray}~~// {#1}}}

\newcounter{algorithmline}[algorithm]
\newcommand{\newalgline}{\refstepcounter{algorithmline}\thealgorithmline} 

\ifx\hidespecial\undefined
	\newcommand{\todo}[1]{ {\color{MyRed}\textbf{TODO:} #1} }

\else
	\newcommand{\todo}[1]{}

\fi

\ifx\hideproof\undefined
	\usepackage{framed}
	\definecolor{leftBarGray}{rgb}{0.8,0.8,0.8}
	\newenvironment{myleftbar}{%
	\MakeFramed {\advance\hsize-\width \FrameRestore}}%
	{\endMakeFramed}
	
\else
	\usepackage{environ}
	\NewEnviron{quoteproof}{}
\fi

\titleformat{\subsubsection}[runin]{\bfseries}{\thesubsubsection~}{0pt}{\bfseries}[:]

\title{A Search Algorithm for Simplicial Complexes}

\author{
Subhrajit Bhattacharya\thanks{
{Department of Mechanical Engineering and Mechanics, Lehigh University, 562 Packard Laboratory, 19 Memorial Drive West, Bethlehem, PA 18015. e-mail: \texttt{sub216@lehigh.edu}}
}
}
\date{\small \monthname, \the\year}

\begin{document}

\maketitle

\begin{abstract}
 We present the `Basic S*' algorithm for computing shortest path through a metric simplicial complex.
 In particular, given a metric graph, $G$, which is constructed as a discrete representation of an underlying configuration space (a larger ``continuous'' space/manifold typically of dimension greater than one), 
 we consider the Rips complex, $\mathcal{R}(G)$, associated with it. Such a complex, and hence shortest paths in it, represent the underlying metric space more closely than what the graph does.
 While discrete graph representations of continuous spaces is convenient for motion planning in configuration spaces of robotic systems, the metric induced in them by the ambient configuration space is significantly different from the metric of the configuration space itself. We remedy this problem using the simplicial complex representation.
 Our algorithm requires only an abstract graph, $G=(V,E)$, and a cost/length function, $d:E\rightarrow \mathbb{R}_+$, as inputs, and no global information such as an embedding or a global coordinate chart is required. The complexity of the Basic S* algorithm is comparable to that of Dijkstra's search, but, as the results presented in this paper demonstrate, the shortest paths obtained using the proposed algorithm represent/approximate the geodesic paths in the original metric space significantly more closely.
\end{abstract}

\section{Introduction}

Computing shortest path in a configuration space is fundamental to motion planning problems in robotics.
While continuous methods for path planning does exist~\cite{Zefran96,ref:Rimon91,ref:Conner03,MAH:VK:07}, 
they suffer from drawbacks, especially in presence of obstacles/holes in the configuration space, such as difficulty in imposing arbitrary optimality criteria (potential/vector field methods~\cite{ref:Rimon91,MAH:VK:07}), large search space (variational methods~\cite{mellingerICRA2012,variationAfflitto10,Desai99motionplanning}), termination at local optimum~\cite{ref:Rimon91,KhoslaVolpeICRA88,KimKhosla02,ref:Conner03} due to non-convex search spaces, and in general lack of rigorous guarantees when the configuration space has an arbitrary topology (non-contractible spaces) or non-trivial geometry (non-convex, general metric spaces).

\begin{wrapfigure}{r}{0.25\columnwidth}
\centering
\vspace{-0.2in}
\fbox{\includegraphics[width=0.25\columnwidth, trim=70 70 5 41, clip=true]{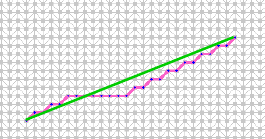}}
\caption{Shortest path in graph is of higher length than the shortest path in the original Euclidean plane.}\label{fig:graph-path-noobs}
\end{wrapfigure}
A robust and popular alternative to the continuous approaches is the discrete approach of graph search-based planning. 
The basic idea behind the approach is to sample points from the configuration space, and construct a graph by 
connecting ``\emph{neighboring}'' vertices with edges (representing actions taking the system from one sampled configuration to another).
Any trajectory in the original configuration space is approximated by a path in the graph~\cite{bondy2007graph}.
One can thus employ any search algorithm like Dijkstra's \cite{dijkstra-dijkstra}, A* \cite{Hart-Astar}, D* \cite{Stentz-dstar}, ARA* \cite{Hansen-anytimeastarjair} or R* \cite{R_star} to search for the optimal path 
in the graph from a start vertex to a goal vertex.
Such a discrete approach for motion planning in graphs are indifferent to the underlying topology/geometry of the configuration space (hence suitable for use in arbitrary configuration spaces), comes with guarantees on algorithmic completeness, termination and optimality in the graph (or bounds on sub-optimality), and are extremely fast.
Such conveniences are precisely the reason that graph search-based approaches have been extremely popular in solving motion planning problems on real robotic systems such as 
motion planning for autonomous vehicles~\cite{Ferguson_2008_6207,Urmson_2008_6189,Montemerlo:2008:JSE:1405647.1405651,UrbanChallenge:VT:09},
planning for robotic arms such as PR2 in cluttered environments~\cite{Ben:PR2:planning},
multi-robotic systems~\cite{conf-icra-SwaminathanPL15,Planning:ICRA:10},
and motion planning for systems involving cables~\cite{cable:separation:IJRR:14,ICRA:14:tethered}.

\vspace{0.1in}
However the major drawback of using such discrete graph-based approaches in motion planning is that the computed paths remain constrained to the graph, which constitutes a small ($1$-dimensional) subset of the original configuration space.
This means that paths that are optimal in the graph need not be optimal in the original configuration space. This issue is typically not remedied by reducing size of the discretization (see Figure~\ref{fig:graph-path-noobs}).
In recent years there have been significant effort in trying to remedy this issue in specific classes of configuration spaces or graphs. All such approaches fall under the general category of what is known as ``\emph{any-angle path planning}'' algorithms~\cite{DBLP:conf-socs-UrasK15}.

The method proposed in this paper, in the same spirit, may be considered as an any-angle path planning algorithm. Instead of planning paths in a graph, we propose an algorithm for finding shortest paths through simplicial complexes. In particular, given a graph, we consider the \emph{Rips complex} of the graph, and compute shortest path in that complex (Figure~\ref{fig:graph-rips-complex}).
The unique features of our proposed method are as follows:
\begin{itemize}[itemsep=0em]
 \item While the input to our algorithm is a metric graph (\emph{i.e}, a graph with specified edge costs/lengths), the underlying structure on which we compute an optimal path is a metric simplicial complex (for a given graph we consider its Rips complex). More generally, our algorithm can be used to compute shortest paths in metric simplicial complexes (not necessarily a Rips complex of a metric graph). 
 \item The input graph can be an arbitrary, abstract metric graph. In particular, we do not require the underlying metric space (whose discrete representation is the graph) to be a subset of flat/Euclidean space (unlike what is required by Theta*~\cite{conf-aaai-NashDKF07,conf-aaai-NashKT10}, ANYA~\cite{Harabor:2013} and Simplicial Dijkstra~\cite{Yershov:Simplicial-dijkstra}). Informally speaking, our method can deal with graphs with ``\emph{non-uniform traversal costs}'' -- both non-homogeneous and anisotropic.
 \item Our method does not require the graph to be embedded in some continuous space or an Euclidean space. In particular, we do not need coordinates for the vertices as input to the algorithm. The only input required to our algorithm is the abstract graph, $G=(V,E)$, and a cost/length function, $d:E\rightarrow \mathbb{R}_+$. Embeddings are constructed locally for simplices as required, and no other data, besides $G$ and $d$, are required. This is in contrast to \cite{Yershov:Simplicial-dijkstra,ferguson2007field}.
  \item Our algorithm is designed for simplicial complexes of arbitrary dimensions and does not require any specific kind of discretization, as long as the simplical complex covers the entire original configuration space (in particular, any arbitrary triangulation of a $2$-dimensional configuration space is sufficient). This, once again, is in contrast to~\cite{ferguson2007field}.
  \item We consider an accurate geometric model in computing the distances, based on local embedding of simplices in a model Euclidean space. This is in contrast to~\cite{Yershov:Simplicial-dijkstra,ferguson2007field}. The accurate model allows us to guarantee that the cost/length of shortest paths computed using the proposed algorithm approaches the true geodesic distance on Riemannian manifolds as the discretization size is made finer.
  \item Our algorithm is local, requiring the abstract graph, $G=(V,E)$, and a cost/length function, $d:E\rightarrow \mathbb{R}_+$, only. No global information such as line-of-sight or global embedding is required.
\end{itemize}


Our focus is the development of an algorithm that can compute optimal paths in arbitrary metric spaces represented by an abstract metric graphs such that the computed path is not restricted to the graph and represents the true geodesic path in the underlying metric space as closely as possible.
Our algorithm is local, requiring only the abstract graph, $G=(V,E)$, and a cost/length function, $d:E\rightarrow \mathbb{R}_+$. 
We use the Dijkstra's search as the backbone for our algorithm, and develop techniques to incorporate simplicial data into it. More efficient versions of the algorithm (incorporating features of heuristic, randomized, incremental and any-time search algorithms) are within the scope of future work.
We believe that in context of robot motion planning, the proposed algorithm is a first, formal use of a finite element method (FEM)~\cite{dhatt2012finite}, where the role of simplicial complexes is well-appreciated.

%
%
%
%

\subsection{Outline of Paper}

In the next sub-section we introduce some preliminary notations and definitions. Following that we introduce the main Basic S* algorithm, the sub-procedures involved in it, and a path reconstruction algorithm. Some theoretical analysis follow. And finally we present simulation results. For better readability, many of the detailed proofs and derivations have been moved to the appendix.

\subsection{Preliminaries}


\begin{figure}[h]
   \centering
     \subfloat[Graph $G = (V,C_1=E)$ (with vertex set $V$ and edge set $E$), and the shortest path in $G$ connecting $\mathsf{a,m}\in V$.]{\fbox{\includegraphics[width=0.25\columnwidth, trim=0 0 0 0, clip=true]{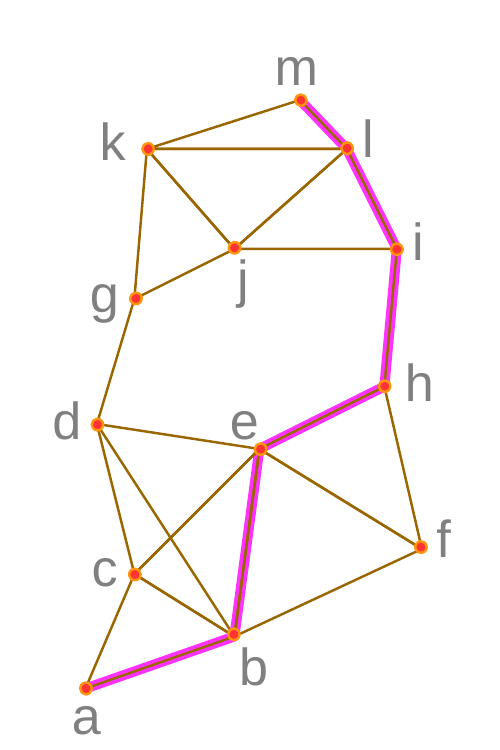}}\label{fig:graph-path}}
     \hspace{0.1in}
     \subfloat[The Rips Complex $\mathcal{R}(G) = \{C_0, C_1, C_2, C_3, \cdots\}$, and the shortest path in (an Euclidean embedding of) $\mathcal{R}(G)$ connecting $\mathsf{a,m}\in V$.]{\fbox{\includegraphics[width=0.25\columnwidth, trim=0 0 0 0, clip=true]{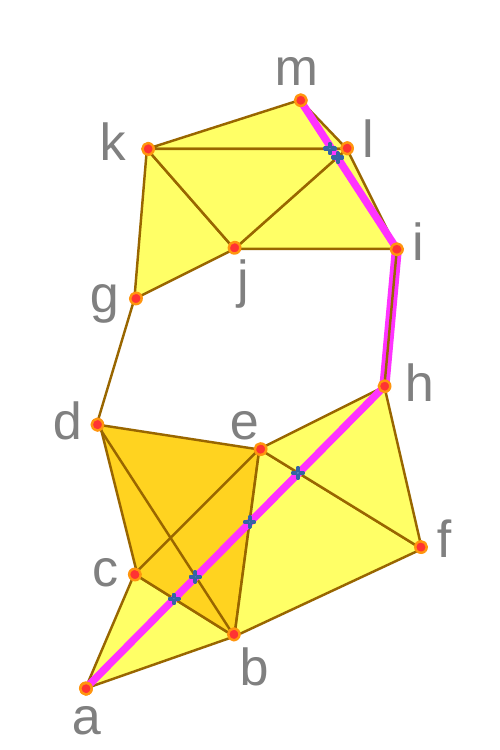}}\label{fig:simplicial-cplx-path}}
  \caption{A graph (left), its Rips complex (right), and shortest paths in those. In this example, $V = \left\{\mathsf{a,b,c,d,e,f,g,h,i,j,k,l,m}\right\}$ is the vertex set, 
  $C_0 = \left\{\mathsf{\{a\},\{b\},\{c\},\{d\},\{e\},\{f\},\{g\},\{h\},\{i\},\{j\},\{k\},\{l\},\{m\}}\right\}$ is the set of $0$-simplices, 
  $C_1 = \big\{\mathsf{ \{a,b\}, \{b,c\}, \{c,a\}, \{c,d\}, \{c,b\}, \{b,d\}, \{b,d\}, \{e,b\}, \{e,c\}, \{b,f\}, \{e,f\}, \{f,h\}, \{e,h\}, \{h,l\}, \{d,g\}, \{g,j\}, \{j,k\}, \{j,l\}, \{k,l\}, \{j,i\},}$ $\mathsf{ \{i,l\}, \{l,m\}, \{k,m\} }\big\}$ is the edge set (the set of $1$-simplices), $C_2 = \big\{\mathsf{ \{a,b,c\}, \{b,c,d\}, \{b,c,e\}, \{e,c,d\}, \{b,e,d\}, \{e,b,f\}, \{f,h,e\},}$ $\mathsf{ \{i,j,l\}, \{j,l,k\}, \{k,l,m\}, \{k,g,j\} }\big\}$ is the set of $2$-simplices, and $C_3 = \big\{ \{b,c,d,e\} \big\}$ is the set of $3$-simplices.}
  \label{fig:graph-rips-complex}
\end{figure}


\begin{definition}[Simplicial Complex~\cite{Hatcher:AlgTop} -- Combinatorial Definition]
 A simplicial complex, $\mathcal{C}$, constructed over a set $V$ (the \emph{vertex set}) is a collection of sets $C_n, ~n=1,2,\cdots$, such that
 \begin{enumerate}[itemsep=0mm]
  \item[i.] An element in $C_n, ~n\geq 0$ is a subset of $V$ and has cardinality $n+1$ (\emph{i.e.}, For all $\sigma \in C_n$, $\sigma \subseteq V, |\sigma| = n+1$). $\sigma$ is called a ``\emph{$n$-simplex}''.
  \item[ii.] If $\sigma \in C_n, ~n\geq 1$, then $\sigma \!-\! v \in C_{n-1}, ~\forall v \in \sigma$. Such a $(n \!-\! 1)$-simplex, $\sigma \!-\! v$, is called a ``\emph{face}'' of the simplex $\sigma$.
 \end{enumerate}
 The simplical complex is the collection $\mathcal{C} = \{C_0, C_1, C_2, \cdots\}$. We also define $C_* = C_0 \cup C_1 \cup C_2 \cup \cdots \subseteq \mathcal{P}(V)$.
\end{definition}

In general, a $n$-simplex is a set containing $n+1$ elements, $\sigma = \{v_0, v_1, v_2, \cdots, v_{n-1}, v_n\}$, where $v_i\in V, ~i=0,1,\cdots,n$.

In algebraic topology, one imparts group or vector space structures on these sets via operation completions, and defines linear maps between those (the \emph{boundary maps}). However, for the purpose of this paper we do not require such algebraic constructions.


\begin{definition}[Rips Complex of a Graph, $\mathcal{R}(G)$]
 If $G$ is an undirected graph with $V$ its vertex set and $E$ its edge set, we define the \emph{Rips Complex of the graph}, $\mathcal{R}(G)$, to be the simplicial complex with an $n$-simplex consisting of every $(n+1)$-tuple of vertices that are all connected to each other (a \emph{clique}). 
 \newline
 In notations, $\mathcal{R}(G) = \{C_0, C_1, C_2, C_3, \cdots\}$ is such that $C_0 = \{\{a\} ~|~ a\in V\}$ and for $\sigma\in C_n, n>1$ and $a,b \in \sigma$, we have $\{a,b\}\in E$. Also, define $C_* = C_0 \cup C_1 \cup C_2 \cup \cdots$.
\end{definition}

The set $C_0$ is the set of $0$-simplices consisting of singleton sets, each containing a single vertex from $V$, while the set $C_1 = E$.
From now on, whenever we refer to a simplicial complex, unless otherwise specified, we will refer to the Rips complex $\mathcal{R}(G) = \{C_0, C_1, C_2, \cdots\}$, for a given graph $G = (C_0, C_1)$.
Figure~\ref{fig:graph-rips-complex} illustrates the Rips complex of a graph with an explicit example of the sets $C_0, C_1, C_2, \cdots$.


\section{Basic S* Algorithm}

\begin{figure}
\centering
\begin{tabular}{ccc}
\subfloat[Vertices $q_1$ and $q_2$ are candidates for being \emph{expanded}.]{\fbox{\includegraphics[width=0.3\textwidth, trim=20 20 20 20, clip=true]{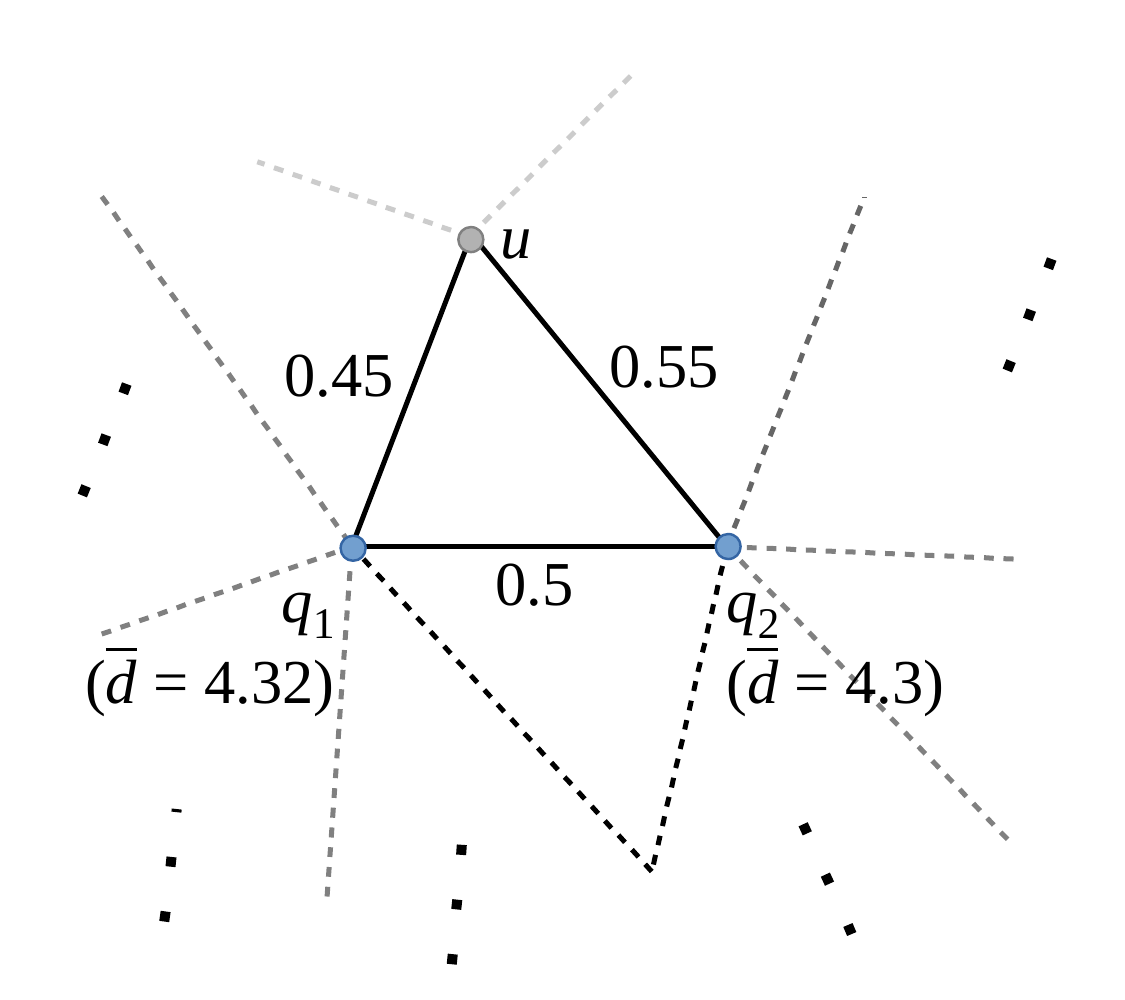}}} & 
\subfloat[\textbf{Expansion of $q_2$ in Dijkstra's or Basic S*:} $q_2$ has a lower $f = \overline{d}$ value (distance from $s$) than $q_1$, hence is expanded first. The $f$-value of $w$ is set to $\overline{d}(q_2) + d(w,q_2) = 4.3+0.55=4.85$ -- the distance to $w$ through the edge $\{q_2,w\}$ (also a $1$-simplex containing $q_2$ and $w$). The ``came-from'' vertex of $w$ is set to $q_2$. This step, in this example, will be same for both Dijkstra's algorithm and Basic S*.]{\fbox{\includegraphics[width=0.3\textwidth, trim=20 20 20 20, clip=true]{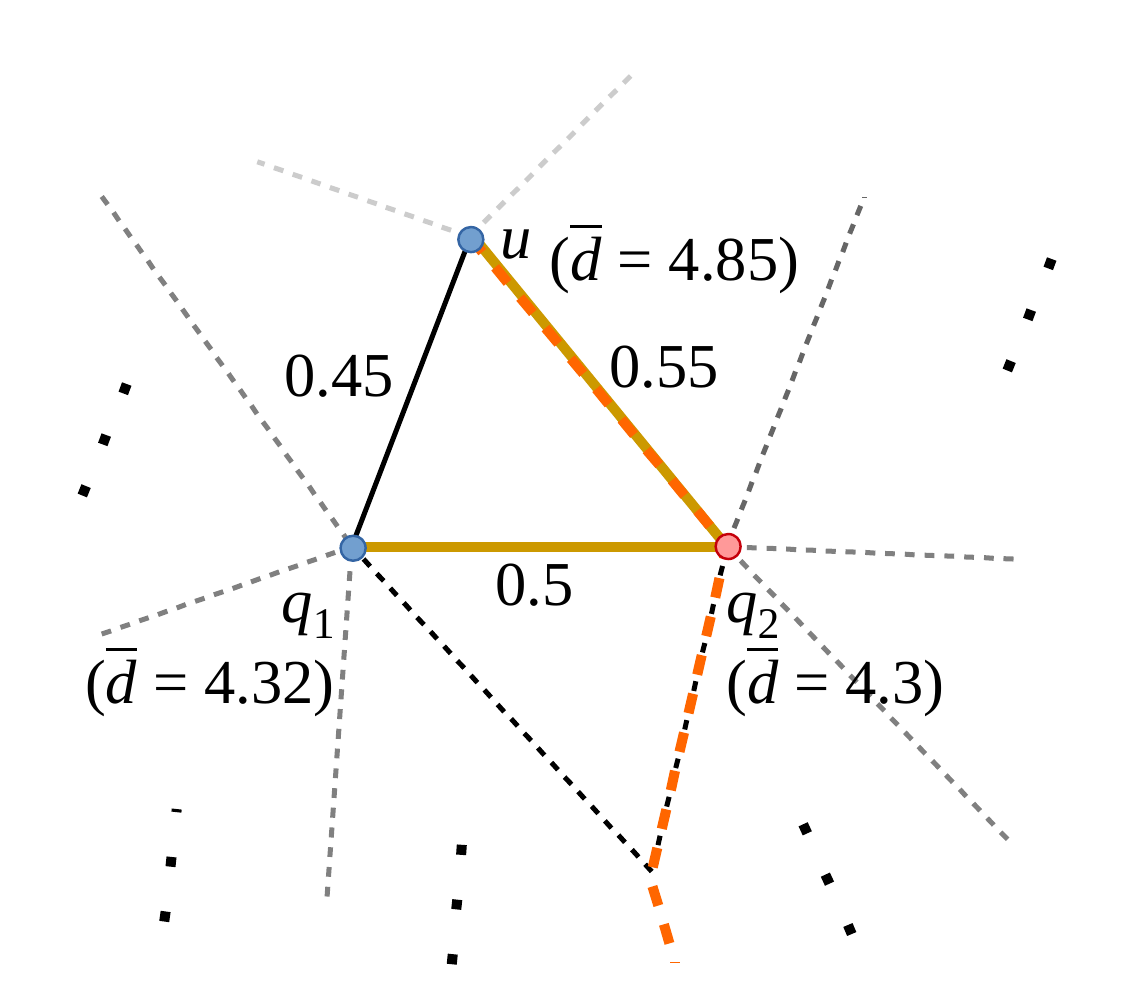}}} & 
\subfloat[\textbf{Expansion of $q_1$ in Dijkstra's:} Expansion of $q_1$ gives a lower $f$-value of $w$ for a path through the edge $\{q_1,w\}$ (since $\overline{d}(q_1) + d(w,q_1) = 4.32+0.45=4.77$), hence its $f$-value is updated, and the ``came-from'' vertex of $w$ is set to $q_1$. This is typical of a search algorithm such as Dijkstra's.]{\fbox{\includegraphics[width=0.3\textwidth, trim=20 20 20 20, clip=true]{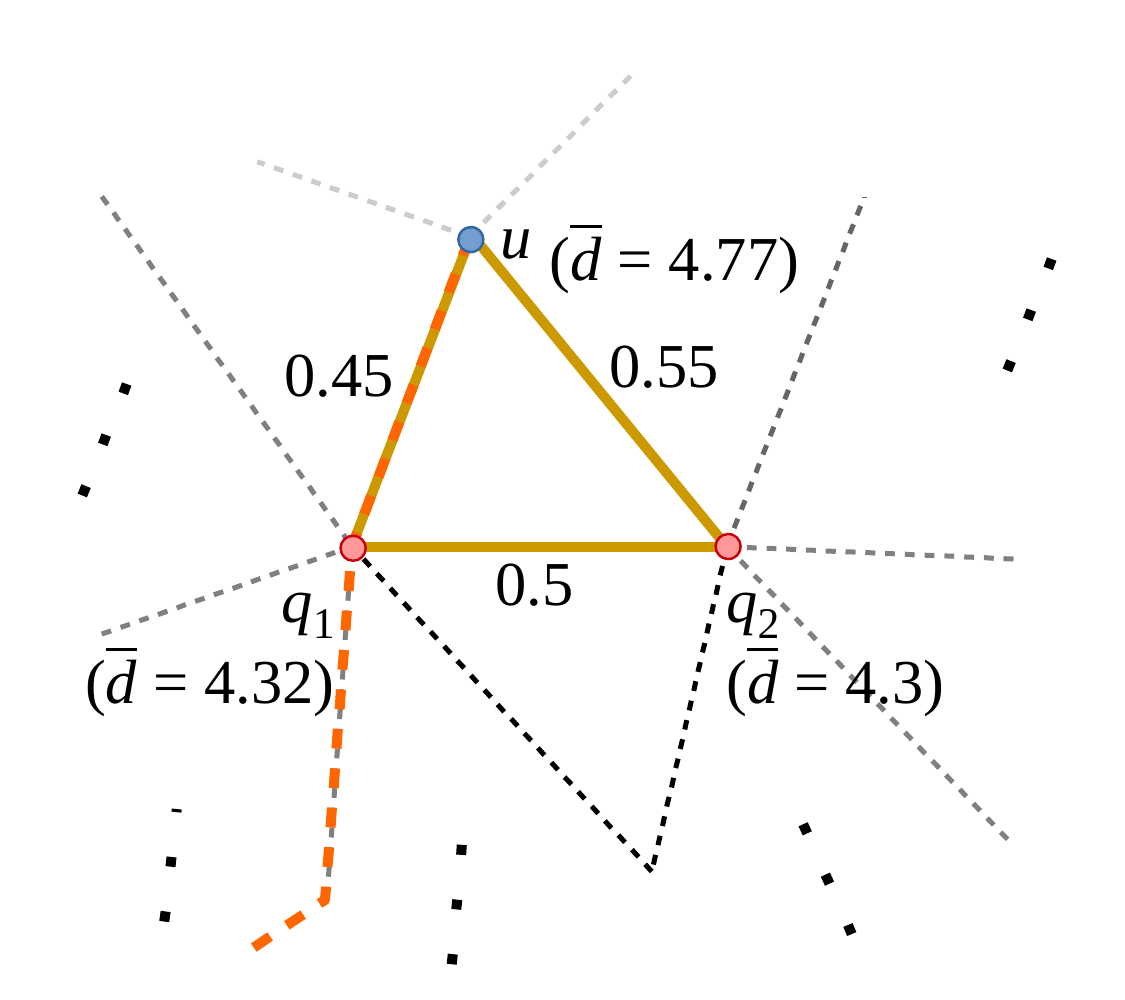}}} \\
%
%
%
\multicolumn{3}{c}{\subfloat[
\textbf{Expansion of $q_1$ in Basic S*:}
\emph{Figure on left:} In Basic S* algorithm, when vertex $q_1$ is expanded, and the $f$-value of $u$ is to be checked for update, 
we construct all the \emph{maximal simplices} attached to $q_1$,
containing $u$, and containing already-expanded vertices.
In this example $\{q_1,q_2,u\}$ is such a maximal $2$-simplex (since $q_2$ has already been expanded). 
The distance to $u$ through this $2$-simplex is computed, and is the candidate for $f$-value update. In this example, if the $f$ values of $q_1$ and $q_2$ are once again $\overline{d}(q_1)=4.32$ and $\overline{d}(q_2)=4.3$, then the distance to $u$ through $\{q_1,q_2,u\}$ is $4.72$, and hence its $f$-value is updated and the ``came-from'' simplex of $u$ is set to $\{q_1,q_2,u\}$.
\emph{Figure on right:} The way we compute the distance through the $2$-simplex involves constructing an embedding of the abstract metric simplex, $\{q_1,q_2,u\}$, in a same-dimensional Euclidean plane and an embedding of the metric $2$-simplex $\{q_1,q_2,o\}$ with the lengths of $q_1o$ and $q_2o$ being equal to the corresponding vertices' $f$-value. Since $\overline{ow}$ intersects $\overline{q_1q_2}$ in the embedding, the length of $\overline{ow}$ gives the desired value of distance to $u$ through $\{q_1,q_2,u\}$.
We refer to this method of computing the distance as \emph{spherical extrapolation}. An alternative to this is \emph{linear extrapolation} (refer to Figure~\ref{fig:d-bar-dist-lin}).]{ \label{fig:simplex-embed}
    \fbox{\includegraphics[width=0.3\textwidth, trim=20 20 20 20, clip=true]{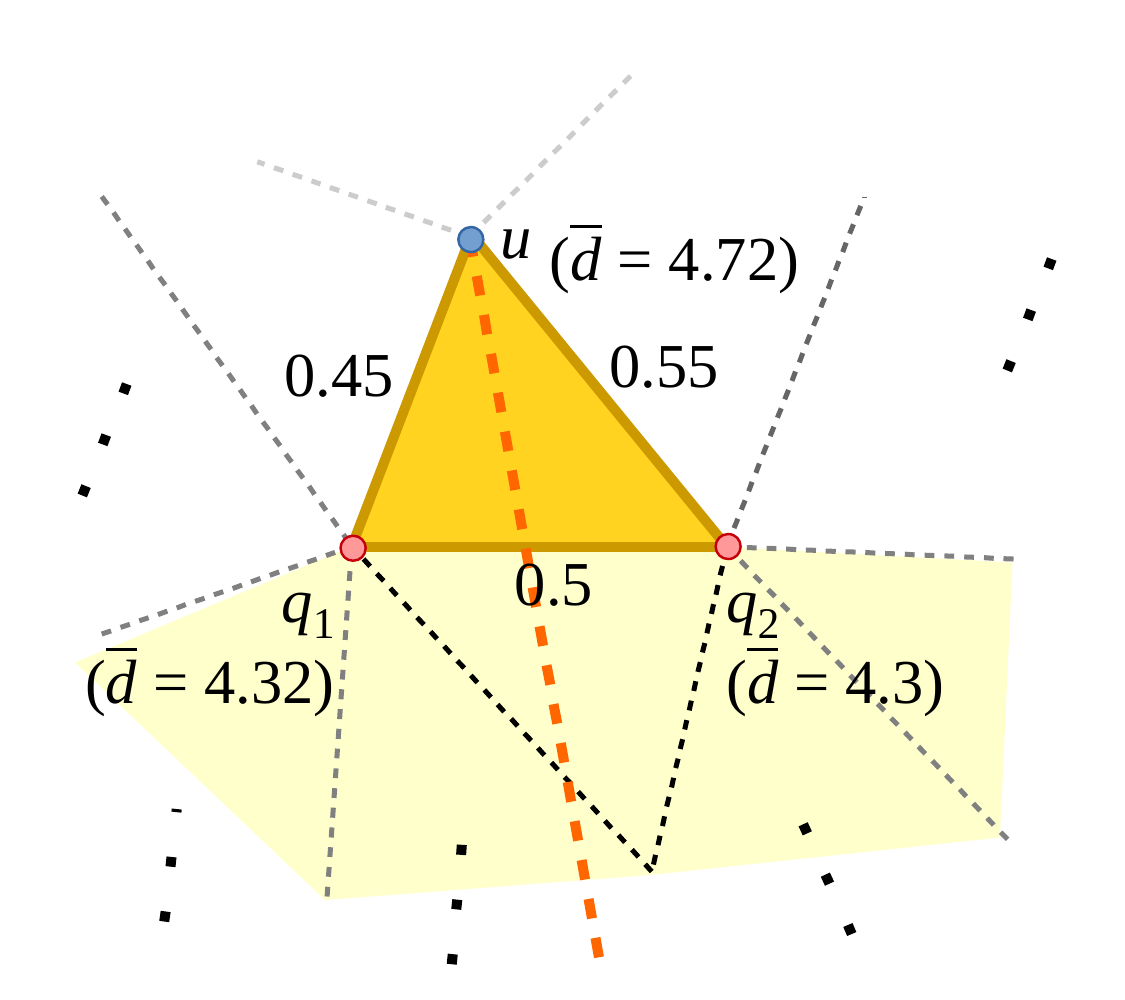}} \hspace{0.2in}
    \fbox{\includegraphics[width=0.56\textwidth, trim=0 5 0 20, clip=true]{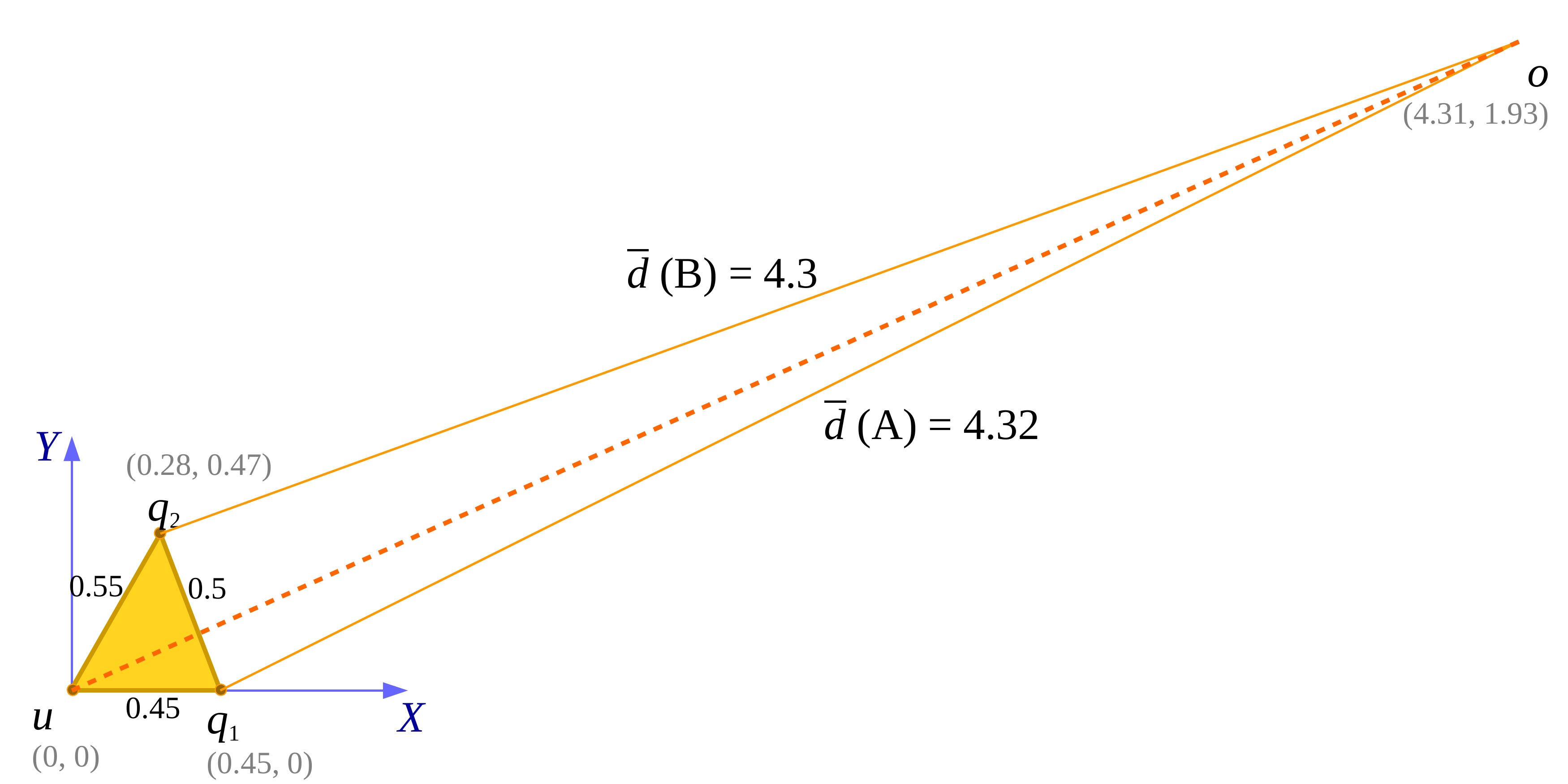}}
    }}
\end{tabular}
\caption{Comparison between Dijkstra's search algorithm for graph, $G$, and the Basic S* search algorithm for $\mathcal{R}(G)$. (a)-(c) shows steps in a typical graph search algorithm. In Basic S*, the steps(a)-(b) are the similar, but step (c) is replaced by the step (d).}
 \label{fig:basic-s-star-simplex}
\end{figure}

The idea behind the S* search algorithm is very similar to the standard search algorithms such as Dijkstra's or A*. However, instead of restricting paths to the graph $G$, it allows paths to pass through simplices of $\mathcal{R}(G)$. Specifically, for a particular vertex, $u\in V$, when updating its
\emph{``came-from''} vertex and 
its minimum distance from the start, 
we don't just replace the earlier values with the new lower value. 
Instead, we construct the maximal simplices (Definition~\ref{def:maximal-simplex}) containing 
the vertex $w$, the expanding vertex, and other already expanded vertices,
and find the shortest path through those simplex. This is illustrated in the example of Figure~\ref{fig:basic-s-star-simplex}.
The complete pseudocode for the \emph{Basic S*} search algorithm (with a single start vertex and no specified goal vertex) is given in Algorithm~\ref{alg:basic-s-star}.

\begin{algorithm}{Basic S*}

 \multicolumn{2}{|l|}{$\left( \overline{d}, c\!f\!p \right)$ = $Basic\_S^*$~$(G,d, s)$} \\
 \multicolumn{2}{|l|}{\!\!\!\!\begin{tabular}{ll} Inputs: &
                                 \!\!\!\!a. Graph $G$ with vertex set $V$ and (undirected) edge set $E=C_1$\\
                               & \!\!\!\!b. A length/cost function on the edge set, $d:E \rightarrow \mathbb{R}_{+}$\\
                               & \!\!\!\!c. Start vertex, $s\in V$.
                              \end{tabular}
                     } \\
 \multicolumn{2}{|l|}{\!\!\!\!\begin{tabular}{ll} Outputs: &
                                 \!\!\!\!a. The distances from $s$ to every vertex in the graph, $\overline{d}: V \rightarrow \mathbb{R}_+$\\
                               & \!\!\!\!b. Came-from-point map, $c\!f\!p: V \rightarrow C_* \times \mathbb{R}^k
                                    $
                              \end{tabular}
                     } \\
 \hline

 \newalgline\label{s-star:first-line} & Initiate $\overline{d}$: Set $\overline{d}(v) := \infty$ for all $v \in V$ \algcomment{distances (implicit declaration for any un-visited vertex).} \\
 \newalgline & Set $\overline{d}(s) \leftarrow 0$ \algcomment{start vertex.} \\
 \newalgline & Initiate $c\!f\!p$: Set ${c\!f\!p} (v) := \emptyset$ for all $v \in V$ \algcomment{came-from point (implicit declaration for any un-visited vertex).} \\
 \newalgline & Set $Q := V$ \algcomment{Set of un-expanded vertices.} \\
 \newalgline\label{s-star:main-loop-start} & \textbf{while} ($Q \neq \emptyset$ ~AND~ stopping criterion not met)  \\
 \newalgline\label{s-star:extract-q} &     \hspace{1.5em} Set $q := {\arg\min}_{q' \in Q} ~~\overline{d}(q')$ \algcomment{Maintained by a heap data-structure.} \\
 \newalgline &     \hspace{1.5em} \textbf{if} $(\overline{d}(q) == \infty)$ \algcomment{cannot reach any other vertex.} \\
 \newalgline &         \hspace{3em} \textbf{break} \\
 \newalgline &     \hspace{1.5em} Set $Q \leftarrow Q - \{q\}$ \algcomment{Remove $q$ from $Q$.} \\
 \newalgline\label{s-star:neighbor-loop} &     \hspace{1.5em} \textbf{for each} ($u \in \mathcal{N}_G(q)$) \algcomment{For each neighbor of $q$ (both expanded and un-expanded).} \\
 \newalgline &         \hspace{3em} Set $S := \{y ~|~ y\in\mathcal{N}_G(q), y\in\mathcal{N}_G(u), y\notin Q\}$ \algcomment{expanded common neighbors of $q$ and $u$.} \\
 \newalgline\label{s-star:attached-maximaal-simplices} &         \hspace{3em} Set $\mathcal{M}\!S := {MaximalSimplices}_{\mathcal{R}(G)} (S;\{u,q\})$ \algcomment{maximal simplices attached to $\{u,q\}$.} \\
 \newalgline\label{s-star:inner-most-loop} &         \hspace{3em} \textbf{for each} ($\sigma\in \mathcal{M}\!S$) \\
 \newalgline\label{s-star:dist-through-simplex} &             \hspace{4.5em} Set $\left(d', (\sigma',\overline{w}') \right) := {DistanceThroughSimplex}_{(d,\overline{d})}(\sigma, u)$ \algcomment{distance to $u$ through simplex $\sigma$.} \\
 \newalgline &             \hspace{4.5em} \textbf{if} ($d' < \overline{d}(u)$) \algcomment{Found better $\overline{d}$-score for this neighbor.} \\
 \newalgline &                 \hspace{6em} Set $\overline{d}(u) \leftarrow d'$ \\
 \newalgline &                 \hspace{6em} Set ${c\!f\!p}(u) \leftarrow (\sigma',\overline{w}')$ \\
 \newalgline &                 \hspace{6em} \textbf{if} $(u \notin Q)$ \algcomment{$u$ is already expanded.} \\
 \newalgline\label{s-star:unexpand} &                 \hspace{7.5em} Set $Q \leftarrow Q \cup \{u\}$. \algcomment{``un-expand'' $u$.} \\
 \newalgline & \textbf{return} $\overline{d}$, ${c\!f\!p}$. \\ 

 \hline
 \label{alg:basic-s-star}
\end{algorithm}

\noindent
Key features of the algorithm:
\begin{enumerate}

 \item Similar to Dijkstra's or A* search, we maintain a list of ``un-expanded'' vertices (\emph{open list}), $Q$, in a heap data structure with the $\overline{d}$-values (which, in search algorithm literature, has be traditionally called $g$-score) of the vertices being the heap keys.
 At each of the while loop iterations (starting on Line~\ref{s-star:main-loop-start}), the un-expanded vertex with lowest $\overline{d}$-value is popped (``expanded'') and the $\overline{d}$-value of its neighbors are checked for improvement (loop starting on Line~\ref{s-star:neighbor-loop}).

 \item Unlike Dijkstra's algorithm, however, we do not restrict this update only to unexpanded neighbors of $q$, and instead check for update for all neighbors, $u$, in $\mathcal{N}_G(q)$. This is because there may be obtuse simplices, that get generated at a later stage, through which the distance to an already expanded vertex may be shorter. This is described in more details in Figure~\ref{fig:obtuse}.

\begin{figure*}[h]
   \centering
     \subfloat[When $q_2$ is expanded, and the $\overline{d}$-value of $q_3$ is updated, the simple $\{q_2,q_3,q_4\}$ has not been generated yet since, besides $q_2$ and $q_3$, there is another un-expanded vertex, $q_4$, that constitutes the simple.]{ \fbox{\includegraphics[width=0.28\textwidth, trim=20 40 20 0, clip=true]{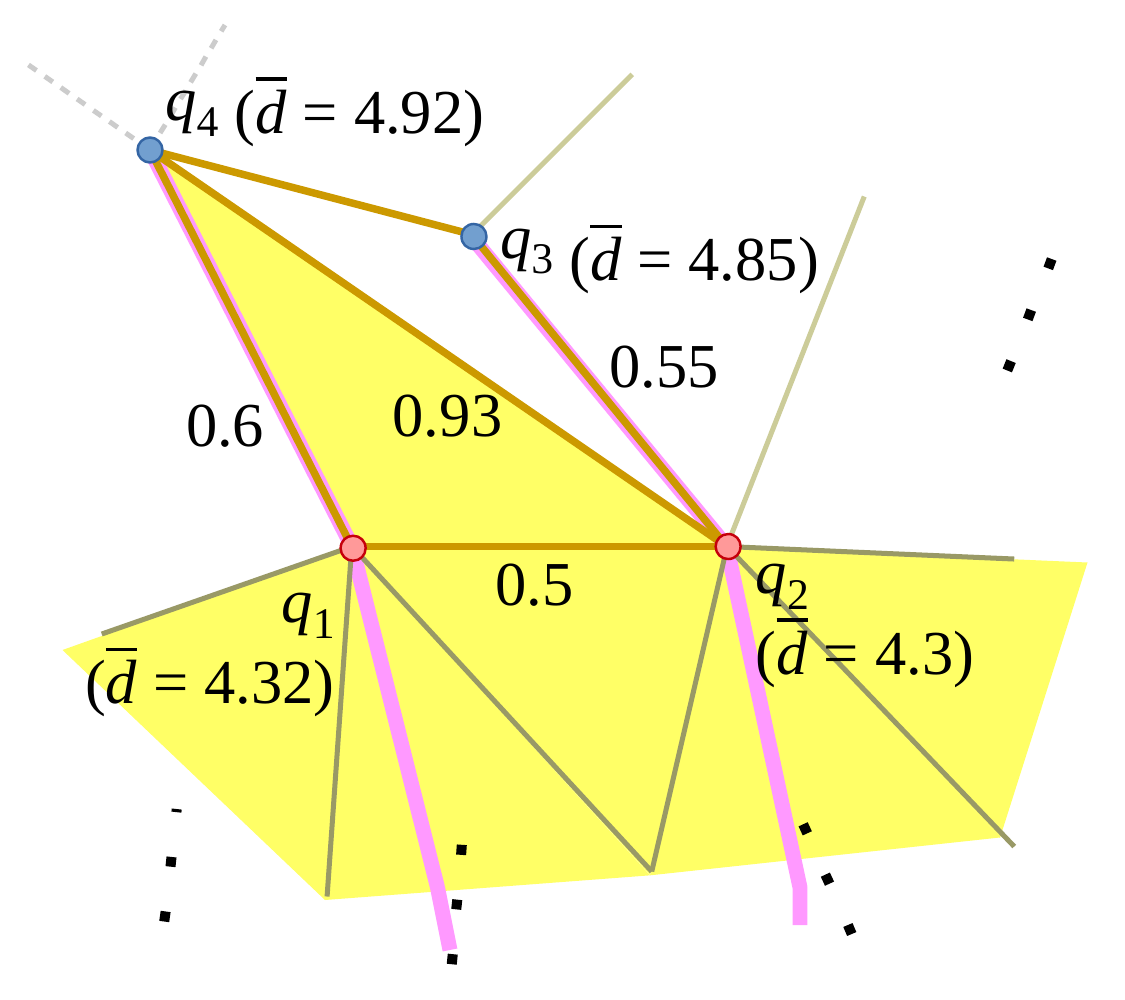}} }
     \hspace{0.02in}
      \subfloat[Next, when $q_3$ is expanded (before $q_4$, since it has a lower $\overline{d}$-value), potential updates are made to the $\overline{d}$-value of $d_4$ and other neighbors of $q_3$, but not $q_3$ itself. Simplex $\{q_2,q_3,q_4\}$ is generated, but only checked for path to $q_4$ (a neighbor of expanding vertex, $q_3$) through that simplex.]{ \fbox{\includegraphics[width=0.28\textwidth, trim=20 40 20 0, clip=true]{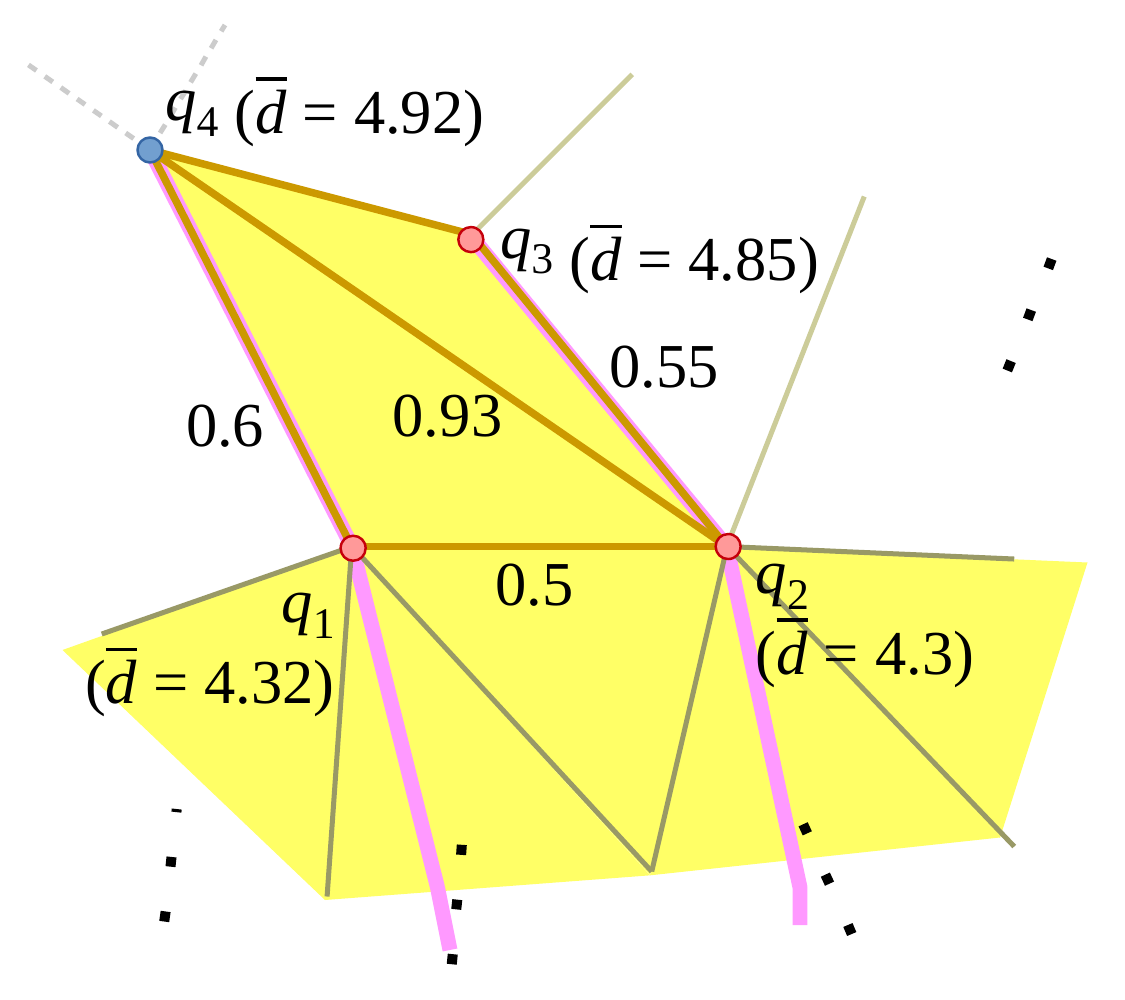}} }
      \hspace{0.02in}
      \subfloat[However, when $q_4$ is being expanded, we can potentially improve the $\overline{d}$-value of its neighbor, $q_3$, for a path through simplex $\{q_2,q_3,q_4\}$, even though $q_3$ has previously been expanded. We chack for improvement, and if the $\overline{d}$-value of a previously expanded vertex is updated, we insert it back to the set $Q$ (a process we call \emph{``un-expanding''} -- Line~\ref{s-star:unexpand} of Algorithm~\ref{alg:basic-s-star}).]{ \fbox{\includegraphics[width=0.28\textwidth, trim=20 40 20 0, clip=true]{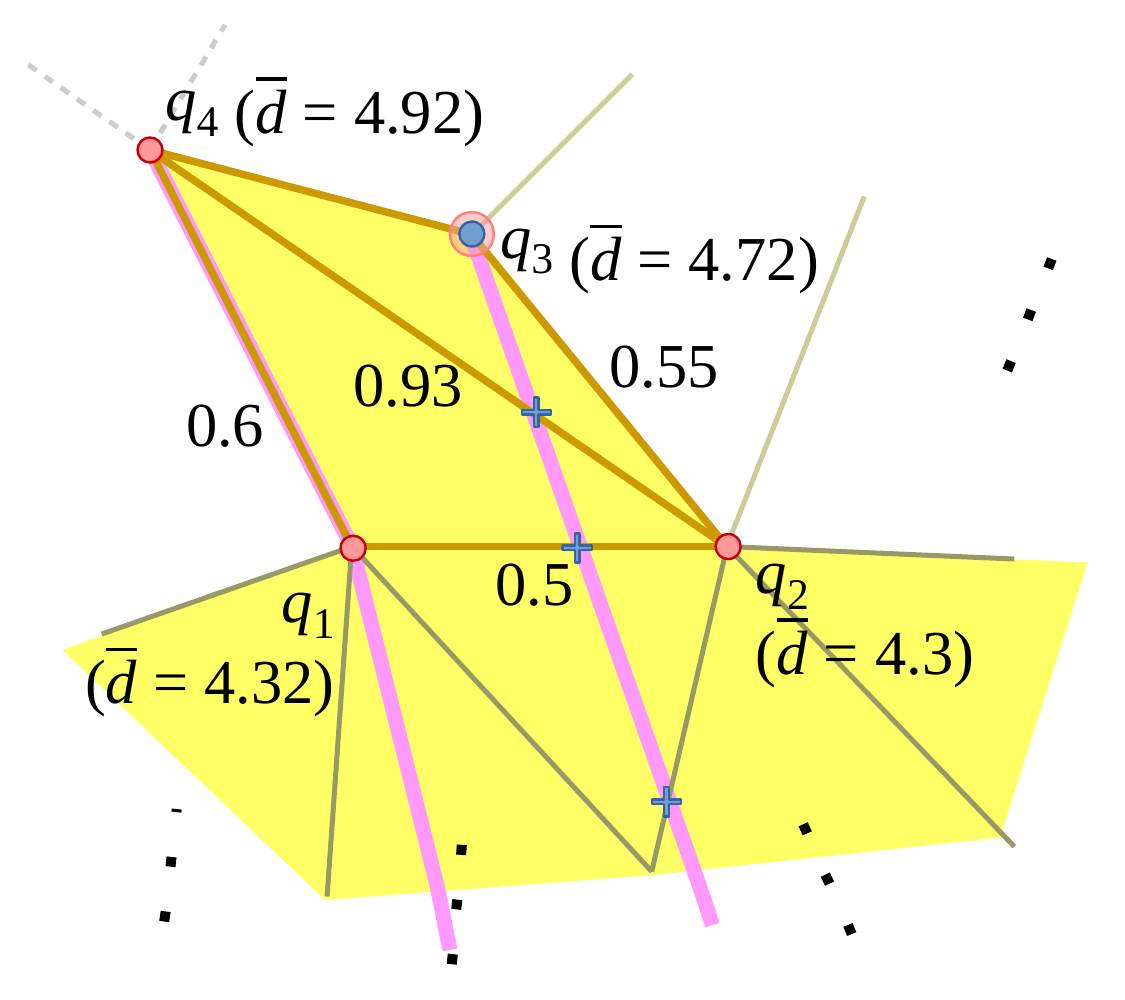}} }
  \caption{Illustration of a scenario where improvement of the $\overline{d}$-value of a vertex, $q_3$, is possible even after it has been expanded. This prompts us to check for potential update of all the neighbors of an expanding vertex rather than only the un-expanded neighbors (Line~\ref{s-star:neighbor-loop} of Algorithm~\ref{alg:basic-s-star}).}
  \label{fig:obtuse}
\end{figure*}
 
 \item When vertex $q$ is being expanded, in order to compute a candidate for updating the $\overline{d}$-value of a neighboring vertex $u$, we generate all maximal simplices (in $\mathcal{R}(G)$) consisting of vertices $q$, $u$ and only other expanded vertices (Line~\ref{s-star:attached-maximaal-simplices}, and procedure `$MaximalSimplices$' is described in more details in Section~\ref{sec:attached-maximal}).

 \item We compute a potential $\overline{d}$-value, $d'$, for updating $u$ for paths through each of these maximal simplices attached to $\{q,w\}$, and choose the lowers out of that as the candidate to test against for update. This is computed by the `$DistanceThroughSimplex$' procedure described in Section~\ref{sec:dist-through-simplex}. This procedure also returns the pair of data, $(\sigma',\overline{w}')$, which represents a ``\emph{came-from point}'' inside simplex $\sigma'$. The ${c\!f\!p}$ is however mostly irrelevant in context of this algorithm.

 \item This is in contrast to the corresponding step in a graph search algorithm, where the potential value to test against is simply the sum of the $\overline{d}$-value at $q$ and the length/cost of the edge $\{q,u\}$.
 
\end{enumerate}


%

\subsection{Computing Attached Maximal Simplices} \label{sec:attached-maximal}

\begin{figure}[h]
   \centering
     \subfloat[In this sub-complex, the set of maximal simplices constructed out of the set $U = \{\mathsf{a,b,c,d,e,f}\}$ is $\mathcal{M}(U) = \big\{\mathsf{ \{a,b,e,c\}, \{a,e,c,d\}, \{ a,c,f\} }\big\}$. This is also the set of maximal simplices attached to $\sigma = \{a,c\}$ and constructed out of $S= \{\mathsf{b,e,d,f}\}$.]{\hspace{0.2in} \fbox{\includegraphics[width=0.25\columnwidth, trim=0 0 0 0, clip=true]{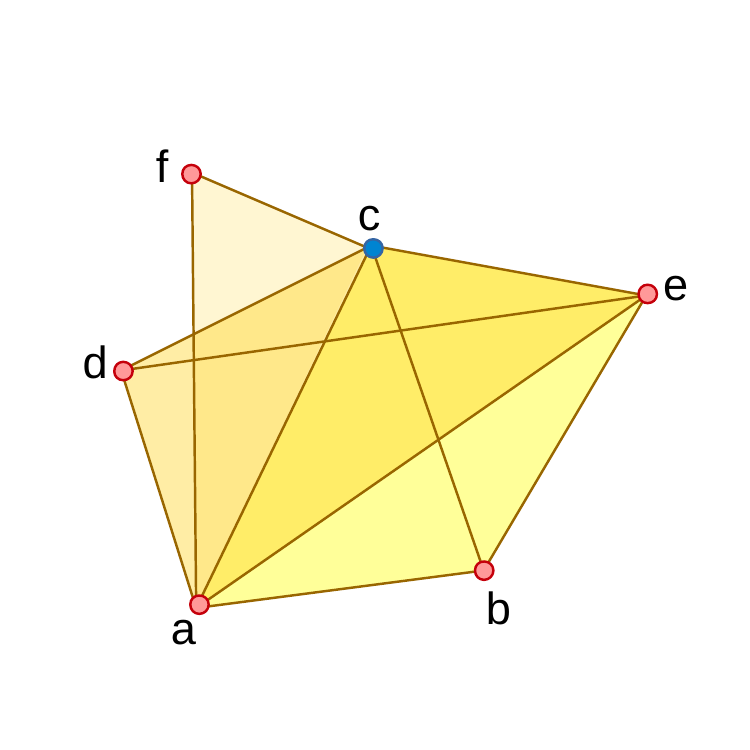}} \hspace{0.2in} \label{fig:maximal-simplex}}
     \hspace{0.1in}
     \subfloat[The set $S= \{\mathsf{b,e,d,f}\}$ is a set neighbors of the $1$-simplex $\sigma = \{a,c\}$. $\mathcal{M}(S) = \big\{\mathsf{ \{d,e\}, \{b,e\}, \{f\} }\big\}$. It is easy to observe that $\mathcal{M}(U) = \{\sigma \cup \chi ~|~ \chi \in \mathcal{M}(S)\}$. Furthermore, since $\mathsf{d}$ and $\mathsf{b}$ are not connected in the complex, $\mathcal{M}(S)$ can be partitioned into subset of simplices containing $\mathsf{d}$ but not $\mathsf{b}$, containing $\mathsf{b}$ but not $\mathsf{d}$, and containing neither.]{\hspace{0.5in} \fbox{\includegraphics[width=0.25\columnwidth, trim=0 0 0 0, clip=true]{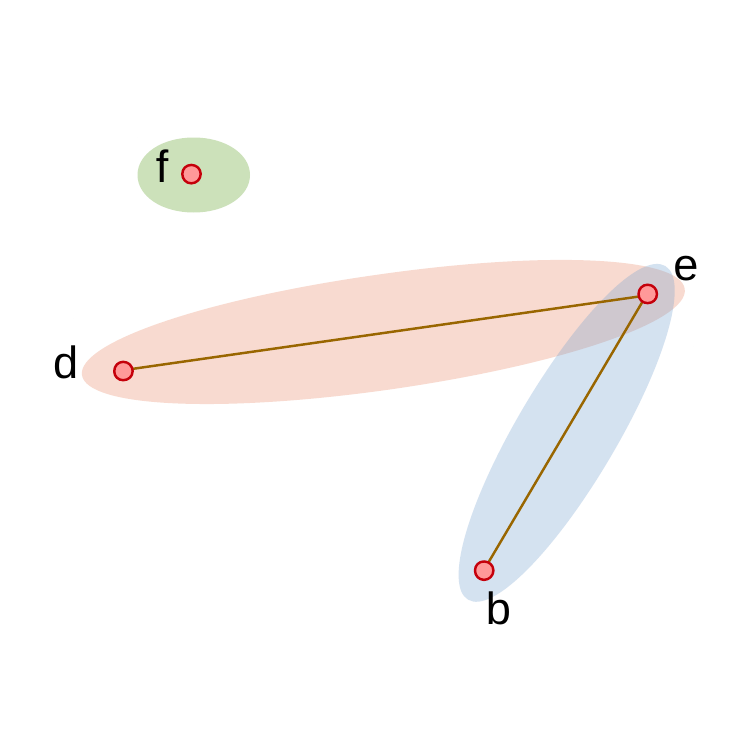}} \hspace{0.5in} \label{fig:maximal-simplex2}}
     \hspace{0.3in}
  \caption{Maximal simplices attached to $\{\mathsf{a,c}\}$ constructed out of its neighbors $\{\mathsf{b,e,d,f}\}$.}
  \label{fig:attached-maximal-simplices}
\end{figure}

\begin{definition}[Maximal Simplices and Maximal Simplices Attached to a Simplex] \label{def:maximal-simplex} ~ 
\begin{enumerate}
 \item[a.] \emph{Maximal Simplex Constructed Out of a Set of Vertices:} A maximal simplex constructed out of a set of vertices, $S\subseteq C_0$, is a subset $\chi\subseteq S$, such that $\chi$ is a simplex in $\mathcal{R}(G)$, and 
 $\chi$ is not a face of any higher-dimensional simplex constituting of the vertices from $S$.
 We refer to the set of maximal simplices created out of $S$ as $\mathcal{M}(S) \subseteq C_*$.
 Formally, $\mathcal{M}(S) = \{\chi ~|~ \chi\subseteq S, \chi\in C_*$ and $\chi \cup a\notin C_* ~\forall a \in S-\chi\}$.
 
 \item[b.] \emph{Neighbors of a Simplex:} A vertex $a\in C_0$ is called a neighbor of a simplex $\sigma\in C_*$ if $\{a,b\}\in C_1$ for all $b\in\sigma$ (\emph{i.e.}, $a$ is connected to every vertex in $\sigma$ by a $1$-simplex). Thus, if $\sigma\in C_n$, then $\{a\} \cup \sigma \in C_{n+1}$. \newline
 The set of all neighbors of $\sigma$ in $G$ is $\mathcal{N}_G(\sigma) = \bigcap_{v\in\sigma} \mathcal{N}_G(v)$ 
 ~(where $\mathcal{N}_G(v)$ is the set of neighbors of $v$ in $G$, excluding $v$ itself).
 
 \item[c.] \emph{Attached Maximal Simplex Constructed Out of a Set of Neighbors:} If $S\subseteq \mathcal{N}_G(\sigma)$ is a set of neighbors of a simplex $\sigma\in C_*$, then the set of maximal simplices constructed out of $S$ and attached to $\sigma$ is the set $\mathcal{M}(S; \sigma) := \mathcal{M}(S \cup \sigma)$. See Figure~\ref{fig:attached-maximal-simplices}.
 \end{enumerate}
\end{definition}

\noindent
The following property is obvious.

\begin{lemma}
 If $\sigma\in C_*$ is a simplex, and 
 $S$ a set of neighbors of $\sigma$,
 (i.e., every vertex in $S$ is connected to every vertex in $\sigma$), then, $\mathcal{M}(\sigma\cup S) = \{\sigma \cup \chi ~|~ \chi \in \mathcal{M}(S)\}$.
\end{lemma}
%


The direct way of computing $\mathcal{M}(S; \sigma)$ will be to check if $\sigma \cup \alpha$ is a simplex in $C_*$ for every $\alpha\in\mathcal{P}(S)$ (the \emph{power set} of $S$). However the complexity of this algorithm would be $O(2^{|S|})$.
The development of a more efficient algorithm for procedure $MaximalSimplices$ relies on the following observation:

\begin{lemma}
 Suppose $\sigma\in C_*$ and $S$ is a set of neighbors of $\sigma$. Identify two vertices, $a,b \in S$ such that $\{a,b\}\notin C_1$ -- {i.e.}, $a$ and $b$ not connected (if such a pair does not exist, then $\sigma\cup S$ is the only maximal simplex).
 Then the set $\mathcal{M}(S; \sigma)$ of maximal simplices constructed out of $S$ and attached to $\sigma$ can be partitioned int three parts:
 \begin{itemize}[itemsep=0mm]
  \item[(i)] Maximal simplices containing $a$, but not containing $b$:~ $\mathcal{M}\left(S \cap \mathcal{N}_G(a);~ \sigma \cup \{a\}\right)$
  \newline (Note: $S \cap \mathcal{N}_G(a)$ does not contain $b$, since $a$ and $b$ are not connected).
  \item[(ii)] Maximal simplices containing $b$, but not containing $a$:~ $\mathcal{M}\left(S \cap \mathcal{N}_G(b);~ \sigma \cup \{b\}\right)$. 
  \item[(iii)] Maximal simplices containing neither $a$ nor $b$:~ $\mathcal{M}(S \!-\! \{a,b\}; \sigma)$.
 \end{itemize}
\end{lemma}

%


\begin{algorithm}{Construct Attached Maximal Simplices}

 \multicolumn{2}{|l|}{$\mathcal{M}\!S$ = ${MaximalSimplices}_{\mathcal{R}(G)}$~$(S;\sigma)$} \\
 \multicolumn{2}{|l|}{\!\!\!\!\begin{tabular}{ll} Inputs: &
                                 \!\!\!\!a. Rips complex, $\mathcal{R}(G) = \{C_0, C_1, C_2, \cdots\}$, of graph $G=(V,E)$.\\ 
                               & \!\!\!\!b. Simplex $\sigma\in C_*$. \\
                               & \!\!\!\!c. A set of neighbors, $S \subseteq \mathcal{N}_G(\sigma)$.
                              \end{tabular}
                     } \\
 \multicolumn{2}{|l|}{\!\!\!\!\begin{tabular}{ll} Outputs: &
                                 \!\!\!\!a. The set of maximal simplices constructed out of $S$ and attached to $\sigma$.
                              \end{tabular}
                     } \\
 \hline
 
 \newalgline & \textbf{if} $(|S| > 1)$ \\
 \newalgline &       \hspace{1.5em}\textbf{for each} $(m \in S, ~n\in S, ~m\neq n)$\\ 
 \newalgline &           \hspace{3em} \textbf{if} $\{m,n\} \notin C_1$ \\
 \newalgline &               \hspace{4.5em} $a := m, ~b := n$ \algcomment{$a,b\in S$ such that they are not connected.} \\
 \newalgline &               \hspace{4.5em} \textbf{break} \\
 \newalgline &  \textbf{if} $a,b$ are undefined \\
 \newalgline &      \hspace{1.5em} $\mathcal{M}\!S := \{\sigma\cup S\}$ \\
 \newalgline &  \textbf{else} \\
 \newalgline &      \hspace{1.5em} $\mathcal{M}\!S_{(i)} := {MaximalSimplices}_{\mathcal{R}(G)} \left(S \cap \mathcal{N}_G(a);~ \sigma \cup \{a\}\right)$ \\
 \newalgline &      \hspace{1.5em} $\mathcal{M}\!S_{(ii)} := {MaximalSimplices}_{\mathcal{R}(G)} \left(S \cap \mathcal{N}_G(b);~ \sigma \cup \{b\}\right)$ \\
 \newalgline &      \hspace{1.5em} $\mathcal{M}\!S_{(iii)} := {MaximalSimplices}_{\mathcal{R}(G)} \left(S - \{a,b\};~ \sigma\right)$ \\
 \newalgline &      \hspace{1.5em} $\mathcal{M}\!S := \mathcal{M}\!S_{(i)} \cup \mathcal{M}\!S_{(ii)} \cup \mathcal{M}\!S_{(ii)}$ \\
 \newalgline & \textbf{return} $\mathcal{M}\!S$.\\
 
 \hline
\label{alg:attached-maximal-simplices}
\end{algorithm}

Factoring in the computational overhead for the search of the pair $\{a,b\}$ and the computation in computing set intersections (which is an $O(k \log k)$ operation for sets of size $k$ maintained using a heap), the complexity of this algorithm is $O(|S|^3 + |S|^2 \log |S|) \sim O(|S|^3)$.


\subsection{Distance of a Vertex Through a Simplex} \label{sec:dist-through-simplex}


\begin{definition}[A Pointed Simplex]
 A \emph{pointed simplex} is a simplex, $\sigma$, with a preferred vertex, $u\in\sigma$, called the \emph{apex} of the simplex. 
\end{definition}

Without loss of generality, we refer to the vertices of a $(n-1)$-simplex, $\sigma$, as ~$v_0, v_1, v_2, \cdots, v_{n-1}$, with $v_0$ being the apex whenever $\sigma$ is pointed, and arbitrarily chosen ordering for $v_1,v_2,\cdots,v_{n-1}$.

\begin{definition}[A Metric Simplex]
 A metric $(n\!-\!1)$-simplex is an $(n\!-\!1)$-simplex, $\sigma$, with a metric defined on the set $\sigma$, $d: \sigma \times \sigma \rightarrow \mathbb{R}_{\geq 0}$, satisfying all the axioms of a metric.
 \newline With $\sigma = \{v_0, v_1, v_2, \cdots, v_{n-1}\}$,
 for brevity we will write $d(v_i, v_j) = d_{i,j} = d_{j,i}$ for all $v_i,v_j\in \sigma$.
Thus, a metric $(n-1)$-simplex is defined by the pair $(\sigma, d)$.
\end{definition}


\begin{definition}[An Euclidean Realizable Metric Simplex]
 A metric $(n\!-\!1)$-simplex, $(\sigma,d)$, is called Euclidean realizable if its constituent vertices can be isometrically embedded in
 an Euclidean space
 (\emph{i.e.}, the Euclidean distance between the embedded vertices are equal to the distances between the vertices in the metric simplex).
\end{definition}

\begin{prop}[Canonical Euclidean Realization of a Metric Simplex] \label{prop:canonical-eu}
 There in an unique embedding of an Euclidean realizable metric $(n\!-\!1)$-simplex, $(\sigma=\{v_0,v_1,\cdots,v_{n-1}\},d)$, 
 given by $e:\sigma \rightarrow \mathbb{R}^{n-1}$ such that
\begin{itemize}[itemsep=0mm]
 \item[i.] The embedded point for the $j^{th}$ vertex has non-zero value for the first $j$ coordinates, with the $j^{th}$ coordinate being non-negative, and zero for the rest. That is, $\mathbf{v}_j := e(v_j) ~=~ [ \vs_{j,0}, ~~\vs_{j,1}, ~\cdots, ~~\vs_{j,j-1}, ~0, ~\cdots,~ 0 ]$, $\vs_{j,j-1}\geq 0, ~~j=0,1,2,\cdots,n-1$.
 \item[i.] $\|\mathbf{v}_i - \mathbf{v}_j\| = d(v_i,v_j) = d_{i,j}$.
\end{itemize}
Explicitly, the embedding can be written using the following recursive formula:
\begin{equation} \label{eq:simplex-eu-embed}
 \vs_{j,k} ~~=~~ \left\{ \begin{array}{ll}
			  { \left( d_{j,0}^2 - d_{j,k+1}^2 + \vs_{k+1,k}^2 + \sum_{p=0}^{k-1} (\vs_{k+1,p}^2 - 2 \vs_{j,p} \vs_{k+1,p}) \right) } ~/~ { 2 \vs_{k+1,k} }, & \text{for $k < j - 1$}, \\
                          \sqrt{d_{j,0}^2 - \sum_{p=0}^{j-2} \vs_{j,p}^2}, & \text{for $k = j - 1$.} \\
                          0, & \text{for $k \geq j$.} \\
                         \end{array} \right.
\end{equation}
where, $\sum_{p=\alpha}^\beta h(p) = 0$ whenever $\beta < \alpha$.
Using \eqref{eq:simplex-eu-embed}, the computation of $\vs_{1,0}, ~\vs_{2,0}, \vs_{2,1}, ~\vs_{3,0}, \vs_{3,1}, \vs_{3,2}, ~\cdots,$ $\vs_{j,0}, \vs_{j,1},\cdots,\vs_{j,j-1}, ~\vs_{j+1,0}, \cdots, \cdots$ can be made in an incremental manner, with the computation of a term in this sequence requiring only the previous terms.

\end{prop}
The proof of this proposition is constructive, and the construction appears in Appendix~\ref{ap:L2-embedding}. An illustration of this embedding of a simple $2$-simplex is shown in Figures \ref{fig:basic-s-star-simplex}\subref{fig:simplex-embed} and \ref{fig:d-bar-dist}.

\begin{definition}[Canonical Euclidean Realization of Metric Simplex]
 The map $e$ described in Proposition~\ref{prop:canonical-eu} is referred to as the \emph{canonical Euclidean realization} of the metric simplex $(\sigma,d)$, and will be referred to as $\mathcal{E}_d (\sigma) : \sigma \rightarrow \mathbb{R}^{n-1}, v_i\mapsto \mathbf{v}_i$.
\end{definition}

\begin{cor}[Embedding Dimension for an Euclidean Realizable Metric Simplex]
 A metric $(n\!-\!1)$-simplex, $(\sigma,d)$, is Euclidean realizable iff its constituent vertices can be isometrically embedded in $\mathbb{R}^{n-1}$.
\end{cor}
The proof of the above corollary follows using rigidity argument and dimension analysis.

\subsubsection{Spherical Extrapolation for Computing Unrestricted $\overline{d}$-distance of Apex}

\begin{prop}
 Given the canonical Euclidean realization, $\mathcal{E}_d(\sigma)=e:v_j \mapsto \mathbf{v}_j = [\vs_{j,0}, ~\vs_{j,1}, \cdots, ~\vs_{j,j-1}, 0, \cdots, 0]$, of a pointed Euclidean realizable metric simplex, $(\sigma,d)$ (with apex $v_0$), and given a map $\overline{d}: \{v_1,v_2,\cdots,v_{n-1}\} \rightarrow \mathbb{R}_{+}$, one can compute a point $\mathbf{o} ~=~   [ \os_{0},  \os_{1},  \cdots, \os_{n-2} ] ~\in \mathbb{R}^{n-1}$, 
 using the following formula:
 \begin{eqnarray} 
  \os_0 & = & \frac{-V+\sqrt{V^2 - 4UW}}{2U}, \nonumber \\
 \os_k  & = & A_k \os_0 ~+~ B_k, \qquad k = 1,2,\cdots,n-2 \label{eq:o-coordinates} 
\end{eqnarray}
 where, 
\begin{equation*}\begin{array}{ll}
 A_1 = \frac{\vs_{1,0} - \vs_{2,0}}{\vs_{2,1}},~~~ & B_1 = \frac{1}{2 \vs_{2,1}}\left( \vs_{2,0}^2 + \vs_{2,1}^2 - \vs_{1,0}^2 +  \overline{d}_1^2 - \overline{d}_{2}^2 \right) \\
 A_k = \frac{\vs_{1,0} - \vs_{k+1,0}}{\vs_{k+1,k}} - \sum_{p=1}^{k-1} \frac{\vs_{k+1,p}}{\vs_{k+1,k}} A_p,~~~ & B_k = \frac{\left( \sum_{p=0}^{k} \vs_{k+1,p}^2 ~-~ \vs_{1,0}^2 +  \overline{d}_1^2 - \overline{d}_{k+1}^2 \right)}{2 \vs_{k+1,k}} - \sum_{p=1}^{k-1} \frac{\vs_{k+1,p}}{\vs_{k+1,k}} B_p, \qquad k \geq 2 \\
 \multicolumn{2}{l}{ U = \left( 1 + \sum_{p=1}^{n-2} A_p^2 \right), ~V = 2 \left( -\vs_{1,0} + \sum_{p=1}^{n-2} A_p B_p \right), ~W = \left( \vs_{1,0}^2 - \overline{d}_1^2 + \sum_{p=1}^{n-2} B_p^2 \right) } \\
 \multicolumn{2}{l}{\overline{d}_j := d(v_j) 
     }
\end{array} 
\end{equation*}
A real solution to \eqref{eq:o-coordinates} exists iff a point $\mathbf{o}$ exists in the same Euclidean space as the embedded metric simplex satisfying $\|\mathbf{o} - \mathbf{v}_j\| = \overline{d}_j := \overline{d}(v_j)$. 
In that case $\mathbf{v}_0$ and $\mathbf{o}$ are points lying on or on the opposite sides of the hyperplane containing $\mathbf{v}_1,\mathbf{v}_2,\cdots,\mathbf{v}_{n-1}$.
\end{prop}
The proof, once again, is constructive, and appears in Appendix~\ref{ap:L2-embedding-o}.

Given an Euclidean realizable metric simplex, $(\sigma,d)$, with apex $v_0$, we can construct its canonical Euclidean realization, $e:=\mathcal{E}_d(\sigma)$, using equations \eqref{eq:simplex-eu-embed}. Also, given the map $\overline{d}$, we can compute the coordinate of $\mathbf{o}$

\begin{definition}[Unrestricted $\overline{d}$-distance of Apex]
 Given the canonical Euclidean realization, $\mathcal{E}_d(\sigma)=e:v_j \mapsto \mathbf{v}_j$, of a pointed Euclidean realizable metric $(n-1)$-simplex, $(\sigma,d)$, with apex $v_0$, 
 we compute
 the point $\mathbf{o}\in\mathbb{R}^{n-1}$ satisfying the given distances $\|\mathbf{o} - e(v_j)\| = \overline{d}_j, ~j=1,2,\cdots,n\!-\!1$ using equations \eqref{eq:o-coordinates}. We thus define the unrestricted $\overline{d}$-distance of $v_0$ to be $\overline{D}^{sph}_{(d,\overline{d})}(\sigma,v_0) = \|\mathbf{o} - e(v_0)\|$.
\end{definition}

\begin{figure}
   \centering
     \subfloat[The $\overline{d}$-distance of apex, $v_0$, through the simplex, $\sigma=\{v_0,v_1,v_2,v_3\}$, is the same as the Euclidean distance between $\mathbf{o}$ and $\mathbf{v}_0$ in the vertices' Euclidean embedding, since $\overline{W}^{sph}_{(d,\overline{d})}(\sigma,v_0) \geq \mathbf{0}$.]{\includegraphics[width=0.3\columnwidth, trim=5 15 20 5, clip=true]{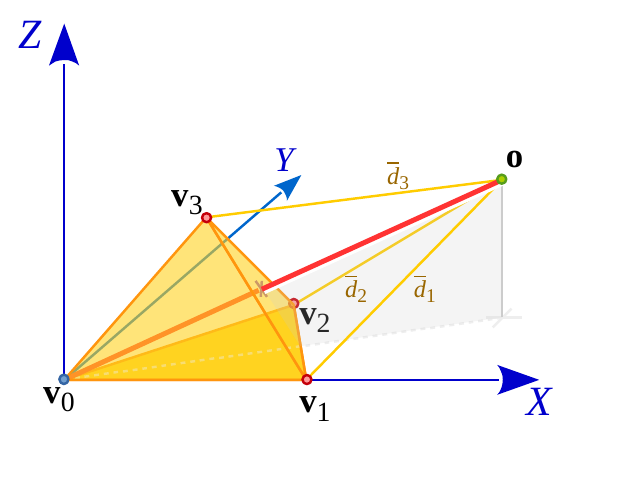} \label{fig:unrestricted-equals-restricted}}
     \hspace{0.1in}
     \subfloat[The line joining $v_0$ and $\mathbf{o}$ intersect $H_0$ outside the triangle formed by $\mathbf{v}_1,\mathbf{v}_2,\mathbf{v}_3$. Not all elements of $\overline{W}^{sph}_{(d,\overline{d})}(\sigma,v_0)$ are non-negative. Thus the $\overline{d}$-distance through simplex is not equal to the unrestricted $\overline{d}$-distance.]{\includegraphics[width=0.3\columnwidth, trim=5 15 20 5, clip=true]{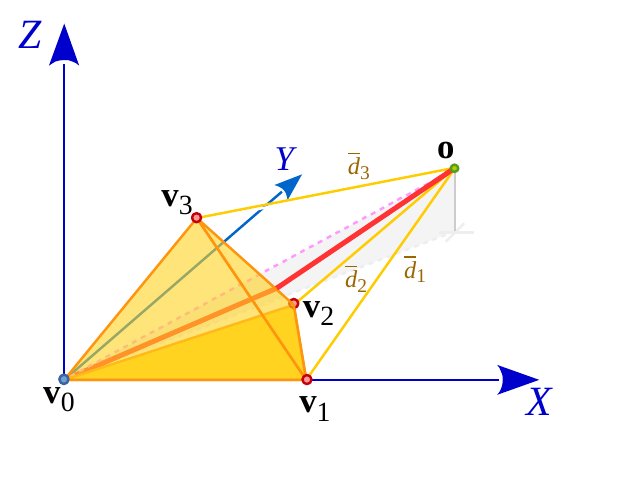} \label{fig:unrestricted-not-equals-restricted}}
     \hspace{0.1in}
      \subfloat[The $\overline{d}$-distance for (b) is computed by computing the unrestricted $\overline{d}$-distance through the face opposite to $\mathbf{v}_1$. This is computed by a completely separate Euclidean realization of the metric simplex $(\{v_0,v_2,v_3\},d)$.]{\includegraphics[width=0.3\columnwidth, trim=5 15 20 5, clip=true]{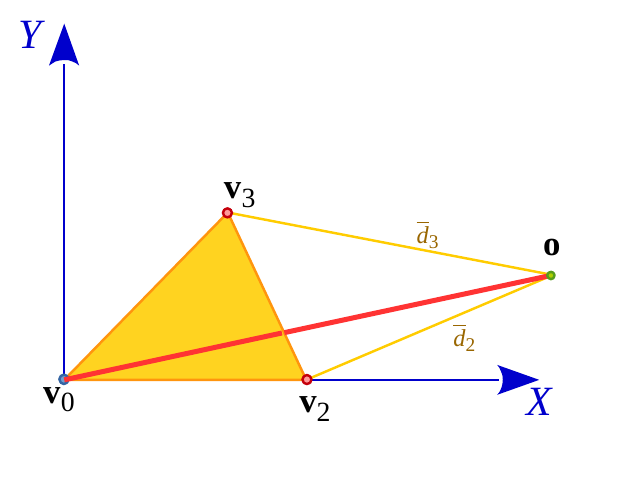} \label{fig:developed}}
  \caption{\emph{Spherical Extrapolation:} Euclidean realization of a metric $3$-simplices, $(\sigma \!=\! \{v_0,v_1,v_2,v_3\}, ~d)$, and given two sets of values for $\overline{d}:\sigma-\{v_0\}\rightarrow \mathbb{R}_+$. $\overline{\mathbf{o}\mathbf{v}_0}$ passes through the face made up of $\mathbf{v}_1,\mathbf{v}_2,\mathbf{v}_3$ in (a), but not in (b). Recursive computation of unrestricted $\overline{d}$-distances trough faces, (c), is required for computed the $\overline{d}$-bar distance for (b).}
  \label{fig:d-bar-dist}
\end{figure}

$\overline{D}^{sph}_{(d,\overline{d})}(\sigma,v_0)$ is the length of the line segment connecting $\mathbf{v}_0$ and $\mathbf{o}$ in the Euclidean realization (Figure~\ref{fig:d-bar-dist}).
This line, $\overline{\mathbf{o}\mathbf{v}_0}$, intersects the hyperplane, $H_0$, containing $\mathbf{v}_1, \mathbf{v}_2,\cdots, \mathbf{v}_{n-1}\}$ at a general point that can be written as $\sum_{i=1}^{n-1} w_i \mathbf{v}_i$, where $\sum_{i=1}^{n-1} w_i = 1$.
The following is a simple geometric consequence, and a derivation appears in Appendix~\ref{ap:L2-embedding-line-plane-intersection}.

\begin{prop}
The point at which the line connecting $\mathbf{v}_0$ and $\mathbf{o}$ intersects the hyperplane $H_0$ is given by $\mathbf{i}_0 = \sum_{i=1}^{n-1} w_i \mathbf{v}_i$, with
\begin{equation} \label{eq:weights-spherical}
 w_k = \frac{{w'}_k}{\sum_{i=1}^{n-1} {w'}_i}, ~~~~k=1,2,\cdots, n\!-\!1.
\end{equation}
where, ${w'}_j$ can be computed recursively using the formula ${w'}_j = \frac{\os_{j-1} - \sum_{i=j+1}^{n-1} {w'}_i \vs_{i,j-1} }{\vs_{j,j-1}}$.
Note that the terms in the sequence ${w'}_{n-1}, {w'}_{n-1}, \cdots, {w'}_1$ can be computed in an incremental manner.

If all the $w_j, ~j=1,2,\cdots,n\!-\!1$ are non-negative, then the line intersects the hyperplane inside (or on the boundary of) the Euclidean realization of the face of the simplex containing $\mathbf{v}_1, \mathbf{v}_2,\cdots, \mathbf{v}_{n-1}$. Otherwise it intersects outside.
\end{prop}

\begin{definition}[Intersection Point in Spherical Extrapolation]
 For the weights computed using equation~\eqref{eq:weights-spherical}, we introduce the map~ $\overline{W}^{sph}_{(d,\overline{d})}(\sigma,v_0) : \sigma\!-\!\{v_0\} \rightarrow \mathbb{R}, ~v_i \mapsto w_i,~ i=1,2,\cdots,n\!-\!1$. 
\end{definition}

The above method of computing unrestricted $\overline{d}$-distance of apex and the weights, $w_j$, relies on the construction of the point $\mathbf{o}$, and identifying the $\overline{d}$-distances as the distances from that point. That's precisely the reason that we refer to this method of computation ``\emph{spherical}''.
In the following sub-section we introduce an alternative to this computation.

\subsubsection{Linear Extrapolation for Computing Unrestricted $\overline{d}$-distance of Apex}

Instead of computing a point, $\mathbf{o}\in\mathbb{R}^{n-1}$, from which the distances of the points $\mathbf{v}_j$ are $\overline{d}_j$, one can compute a $(n\!-\!2)$-dimensional hyperplane, $I$, from which the distances of the points $\mathbf{v}_j$ are $\overline{d}_j$. This gives us an alternative way of computing the $\overline{d}$-distance of the apex, $v_0$ (Figure~\ref{fig:d-bar-dist-lin}). The following proposition summarizes the computation, the proof of which appears in Appendix~\ref{ap:dist-from-plane}.
\begin{prop}
 Given the canonical Euclidean realization, $\mathcal{E}_d(\sigma)=e:v_j \mapsto \mathbf{v}_j = [\vs_{j,0}, ~\vs_{j,1}, \cdots, ~\vs_{j,j-1}, 0, \cdots, 0]$, of a pointed Euclidean realizable metric simplex, $(\sigma,d)$ (with apex $v_0$), and given a map $\overline{d}: \{v_1,v_2,\cdots,v_{n-1}\} \rightarrow \mathbb{R}_{+}$, one can compute a hyperplane, $I$, described by the equation
 $\mathbf{u}\cdot \mathbf{x} + \mu = 0$, where, $\mathbf{x}\in\mathbb{R}^{n-1}$ is a point on the hyperplane, $\mathbf{u} = [\us_{0}, \us_{1}, \cdots, \us_{n-2}] \in\mathbb{R}^{n-1}$ is an unit vector orthogonal to the plane, and $\mu$ is a constant
 using the following formulae:
 \begin{eqnarray} \label{eq:I-plane}
\mu & = & \frac{-Q + \sqrt{Q^2 - 4PR}}{2P}, \nonumber \\
\us_{k} & = & M_{k}~\mu + N_{k}, \qquad k=0,1,2,\cdots,n\!-\!2
\end{eqnarray}
 where, 
\begin{equation*}\begin{array}{ll}
M_0 = -1,~~~ & N_0 = \overline{d}_1 \\
 M_k = -\left( 1 + \frac{1}{\vs_{k+1,k}} \sum_{p=0}^{k-1} M_p \vs_{k+1,p} \right),~~~ & N_k = \left( \overline{d}_{k+1} - \frac{1}{\vs_{k+1,k}}\sum_{p=0}^{k-1} N_p \vs_{k+1,p}  \right), \qquad k \geq 1 \\
 \multicolumn{2}{l}{ P = \sum_{j=0}^{n-1} M_j^2, \qquad Q = 2 \sum_{j=0}^{n-1} M_j N_j, \qquad R = \sum_{j=0}^{n-1} N_j^2 - 1 } \\
 \multicolumn{2}{l}{\overline{d}_j := d(v_j) 
     }
\end{array} 
\end{equation*}
A real solution to \eqref{eq:I-plane} exists iff a plane $I$ exists in the same Euclidean space as the embedded metric simplex 
such that the distances of the points $\mathbf{v}_j$ from the plane $I: \mathbf{u}\cdot \mathbf{x} + \mu = 0$ are $\overline{d}_j$ for $j=1,2,\cdots,n-1$.
In that case
the projection of the simplex constituting of points $\mathbf{v}_1,\mathbf{v}_2,\cdots,\mathbf{v}_{n-1}$ on to the hyperplane $I$, 
and the point $\mathbf{v}_0 = \mathbf{0}$
lie on the opposite sides of the hyperplane containing $\mathbf{v}_1,\mathbf{v}_2,\cdots,\mathbf{v}_{n-1}$.
\end{prop}

\begin{definition}[Unrestricted $\overline{d}$-distance of Apex]
 Given the canonical Euclidean realization, $\mathcal{E}_d(\sigma)=e:v_j \mapsto \mathbf{v}_j$, of a pointed Euclidean realizable metric $(n-1)$-simplex, $(\sigma,d)$, with apex $v_0$, 
 we compute
 the plane, $I$, described by the equation $\mathbf{u}\cdot \mathbf{x} + \mu = 0$ (with $\mathbf{u}$ an unit vector)
 satisfying 
 $\mathbf{u}\cdot \mathbf{v}_j + \mu = \overline{d}_j, ~j=1,2,\cdots,n\!-\!1$,
 using equations \eqref{eq:I-plane}. 
 We thus define the unrestricted $\overline{d}$-distance of $v_0$ to be $\overline{D}^{lin}_{(d,\overline{d})}(\sigma,v_0) = \mathbf{u}\cdot \mathbf{v}_0 + \mu$.
\end{definition}

\begin{figure}
   \centering
     \subfloat[The $\overline{d}$-distance of apex, $v_0$, through the simplex, $\sigma=\{v_0,v_1,v_2,v_3\}$, is the same as the Euclidean distance between $\mathbf{v}_0$ and the plane $I$ in the vertices' Euclidean embedding, since $\overline{W}^{lin}_{(d,\overline{d})}(\sigma,v_0) \geq \mathbf{0}$.]{\includegraphics[width=0.3\columnwidth, trim=5 15 20 5, clip=true]{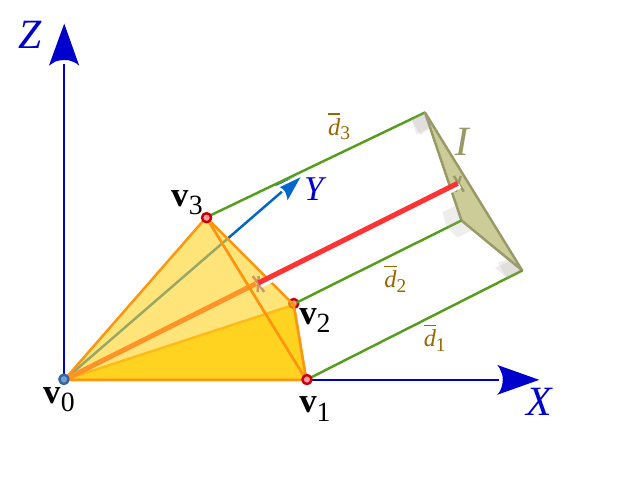} \label{fig:d-bar-dist-lin-a}}
     \hspace{0.1in}
     \subfloat[The perpendicular dropped from $v_0$ to $I$ intersect $H_0$ outside the triangle formed by $\mathbf{v}_1,\mathbf{v}_2,\mathbf{v}_3$. Not all elements of $\overline{W}^{lin}_{(d,\overline{d})}(\sigma,v_0)$ are non-negative. Thus the $\overline{d}$-distance through simplex is not equal to the unrestricted $\overline{d}$-distance.]{\includegraphics[width=0.3\columnwidth, trim=5 15 20 5, clip=true]{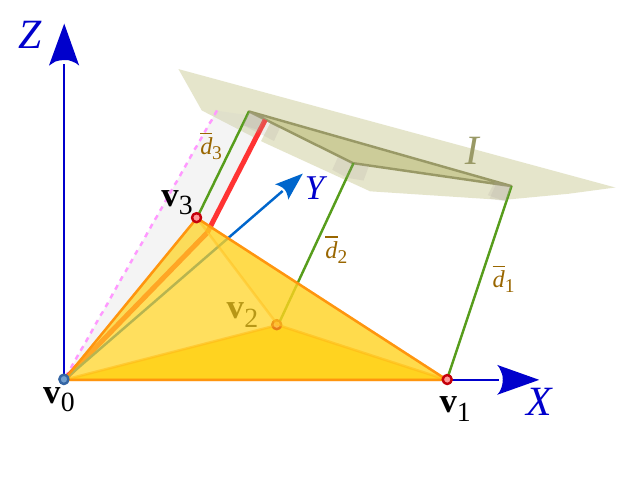} \label{fig:d-bar-dist-lin-b}}
  \caption{\emph{Linear Extrapolation:} Euclidean realization of a metric $3$-simplices, $(\sigma, ~d)$, and given two sets of values for $\overline{d}:\sigma-\{v_0\}\rightarrow \mathbb{R}_+$. The perpendicular dropped from $\mathbf{v}_0$ to $I$ passes through the face made up of $\mathbf{v}_1,\mathbf{v}_2,\mathbf{v}_3$ in (a), but not in (b).
  }
  \label{fig:d-bar-dist-lin}
\end{figure}


\begin{prop}
The point at which the perpendicular dropped from $\mathbf{v}_0$ to the plane, $I$, intersects the hyperplane $H_0$ is given by $\mathbf{i}_0 = \sum_{i=1}^{n-1} w_i \mathbf{v}_i$, with
\begin{equation} \label{eq:weights-linear}
 w_k = \frac{{w'}_k}{\sum_{i=1}^{n-1} {w'}_i}, ~~~~k=1,2,\cdots, n\!-\!1.
\end{equation}
where, ${w'}_j$ can be computed recursively using the formula ${w'}_j = \frac{\us_{j-1} - \sum_{i=j+1}^{n-1} {w'}_i \vs_{i,j-1} }{\vs_{j,j-1}}$.
Note that the terms in the sequence ${w'}_{n-1}, {w'}_{n-1}, \cdots, {w'}_1$ can be computed in an incremental manner.

If all the $w_j, ~j=1,2,\cdots,n\!-\!1$ are non-negative, then the line intersects the hyperplane inside (or on the boundary of) the Euclidean realization of the face of the simplex containing $\mathbf{v}_1, \mathbf{v}_2,\cdots, \mathbf{v}_{n-1}$. Otherwise it intersects outside.
\end{prop}

\begin{definition}[Intersection Point in Linear Extrapolation]
 For the weights computed using equation~\eqref{eq:weights-linear}, we introduce the map~ 
$\overline{W}^{lin}_{(d,\overline{d})}(\sigma,v_0) : \sigma\!-\!\{v_0\} \rightarrow \mathbb{R}, ~v_i \mapsto w_i, ~i=1,2,\cdots,n\!-\!1$.
\end{definition}

A feature of the linear method is that the $\overline{d}$-distances being distances from a plane, there is no notion of triangle inequality or other properties of a metric that can be defined on $\overline{d}$. This may or may not be desirable under different situations.
Given a fine-enough discretization, both the spherical and linear approaches should approximate the underlying metric relatively well.
In our implementation we use only the spherical extrapolation.

\subsection{Algorithm for Computing $\overline{d}$-distance Through Simplex}

As discussed, if some of the weights given by $\overline{W}^{*}_{(d,\overline{d})}(\sigma,v_0)$ are negative, then the line of the shortest length from the apex to $\mathbf{o}$ (in spherical extrapolation) or $I$ (in linear extrapolation) will not 
pass through the face simplex constituting of $\mathbf{v}_1, \mathbf{v}_2,\cdots, \mathbf{v}_{n-1}$. In that case we need to compute the length of the shortest path that intersects $H_0$ inside the simplex constituting of $\mathbf{v}_1, \mathbf{v}_2,\cdots, \mathbf{v}_{n-1}$.
Such a path, clearly, will pass through one of the faces opposite to $\mathbf{v}_1, \mathbf{v}_2, \cdots$ or $\mathbf{v}_{n-1}$ and is the $\overline{d}$-distance of $v_0$ restricted to the face (see Figure~\ref{fig:d-bar-dist}\subref{fig:unrestricted-not-equals-restricted} or \ref{fig:d-bar-dist-lin}\subref{fig:d-bar-dist-lin-b}).
Thus, our algorithm for computing $\overline{d}$-distance through the simplex is one that computes and compares the $\overline{d}$-distance of $v_0$ restricted to the faces until a set of non-negative weights are found. The pseudocode is given below.

%
%
%

\begin{algorithm}{Compute Distance Through Simplex}

 \multicolumn{2}{|l|}{$\left( d', (\sigma',\overline{w}') \right)$ = ${DistanceThroughSimplex}_{(d,\overline{d})}$~$(\sigma,u)$} \\
 \multicolumn{2}{|l|}{\!\!\!\!\begin{tabular}{ll} Inputs: &
                                 \!\!\!\!a. The metric simplex, $\sigma$, with distance between vertices $d:\sigma\times\sigma\rightarrow \mathbb{R}_+$.\\ 
                               & \!\!\!\!b. Apex, $u\in\sigma$. (Thus, letting $\sigma = \{v_0\!:=\!u, v_1, v_2, \cdots, v_{n\!-\!1}\}$.)  \\
                               & \!\!\!\!c. $\overline{d}$-values, $\overline{d}: \sigma\!-\!\{u\} \rightarrow \mathbb{R}_+$.
                              \end{tabular}
                     } \\
 \multicolumn{2}{|l|}{\!\!\!\!\begin{tabular}{ll} Outputs: &
                                 \!\!\!\!a. $\overline{d}$-distance to $u$ through simplex $\sigma$. \\
                               & \!\!\!\!b. Came-from point, $(\sigma',\overline{w}')$, on a subsimplex $\sigma' \subseteq \sigma\!-\!\{u\}$.
                              \end{tabular}
                     } \\
 \hline
 
 \newalgline & If $(\sigma, u)$ exists in lookup table with the specified $\overline{d}$-values for $\sigma\!-\!\{u\}$, \\ & \qquad return its distance-through-simplex value and came-from point. \\
 \newalgline & $\overline{w} := \overline{W}^{*}_{(d,\overline{d})}(\sigma,u)$ \algcomment{weights associated with $v_i, i=1,2\cdots,n\!-\!1$} \\
 \newalgline & \textbf{if} $\overline{w}(v_i) \geq 0, ~\forall i=1,2,\cdots,n\!-\!1$ \algcomment{all weights are non-negative} \\
 \newalgline &     \hspace{1.5em} $d' := \overline{D}^{*}_{(d,\overline{d})}(\sigma,u)$. \\
  \newalgline &    \hspace{1.5em} $\sigma' := \{v \in \sigma\!-\!\{u\} ~|~ \overline{w}(u) > 0 \}$ \algcomment{simplex constituting of vertices with non-zero weights} \\
  \newalgline &    \hspace{1.5em} $\overline{w}'(v) := \overline{w}(v), ~~\forall v\in\sigma'$ \algcomment{weights describing \emph{came-form} point in simplex $\sigma'$} \\
 \newalgline & \textbf{else} \\
 \newalgline &    \hspace{1.5em} $d' := \infty$ \\
  \newalgline &     \hspace{1.5em} \textbf{foreach} $(i \in \{1,2,\cdots,n\!-\!1\})$ \\
  \newalgline &         \hspace{3em} \textbf{if} $(\overline{w}(v_i) < 0)$ \algcomment{Need to check $\overline{d}$-value through face.} \\ 
  \newalgline &             \hspace{4.5em} $\left(d'',(\sigma'',\overline{w}'')\right) = {DistanceThroughSimplex}_{(d,\overline{d})} ~(\sigma\!-\!v_i, ~u)$ \algcomment{distance through face opposite to $v_i$.} \\
  \newalgline &             \hspace{4.5em} Insert an entry for $(\sigma\!-\!v_i, u)$ with the specified $\overline{d}$-values for $\sigma\!-\!\{v_i,u\}$ into lookup table, \\ & \hspace{4.5em} \qquad with $d'$ its distance-through-simplex value and $(\sigma'',\overline{w}'')$ its came-from point. \\
  \newalgline &             \hspace{4.5em} \textbf{if} $d'' < d'$ \\
  \newalgline &                 \hspace{6em} $d' \leftarrow d''$ \\
    \newalgline &                 \hspace{6em} $(\sigma',\overline{w}') \leftarrow (\sigma'',\overline{w}'')$ \algcomment{came-from point.} \\
  \newalgline & \textbf{return} $d', (\sigma',\overline{w}')$ \\
  
 \hline
\label{alg:dist-through-simplex}
\end{algorithm}

\noindent
where, the ``$*$'' in $\overline{W}^{*}$ and $\overline{D}^{*}$ refers to either ``$sph$'' or ``$lin$'' depending on the chosen extrapolation method.

Note that the procedure also return the pair $(\sigma',\overline{w}')$, where $\sigma'$ is a simplex (consisting of vertices with non-zero weights) and $\overline{w}': \sigma' \rightarrow (0,1]$ is a map that associates weights to the vertices in the simplex.
These two pieces of information, $(\sigma',\overline{w}')$, together describes a point inside the simplex $\sigma'$ as the convex combination $\sum_{v\in\sigma'}^m \overline{w}(v) ~v$.


During the recursive calls to $DistanceThroughSimplex$, it is possible that the procedure is called on the same simplex multiple times (by different faces). In order to avoid re-computation of the $\overline{d}$-value through the same simplex multiple times, we maintain a lookup table of the simplices (as a hash table maintained globally across all calls to $DistanceThroughSimplex$), and return the $DistanceThroughSimplex$ value if it exists. The entries in the table (the \emph{hash key}) are distinguished by the simplex vertices (unordered), the simplex's apex as well as the $\overline{d}$-value of the other vertices.


\section{Path Reconstruction}

As described, abstract point inside a simplicial complex can be described by two pieces of information: \emph{i.} a $m$-simplex, $\rho=\{\nu_0,\nu_1,\cdots,\nu_m\}\in C_*$, inside (or on the boundary of) which the point lies, and \emph{ii.} a set of positive weights associated with each vertex such that they add up to $1$ (we represent the weights by the map $\overline{w}: \sigma \rightarrow \mathbb{R}_{\geq 0}$,
with $\sum_{v\in\rho} \overline{w}(v) = \sum_{i=0}^{m} \overline{w}(\nu_i) = 1$).
The point itself would then be the abstract convex combination $\sum_{v\in\rho} \overline{w}(v) = \sum_{i=0}^m \overline{w}(v_i) ~v_i$.
The pair $p = (\sigma, \overline{w})$ described a point, and
a (piece-wise linear) path in a simplicial complex can thus be described as a sequence of points $p_i = (\sigma_i, \overline{w}_i), ~{i=0,1,2,\cdots}$

The primary output of the Basic S* algorithm (Algorithm~\ref{alg:basic-s-star}) is a $\overline{d}$-value for every vertex in $G$.
In order to find the shortest path connecting $s\in V$ with some arbitrary $g\in V$ (which has been expanded), like any search algorithm, we need to reconstruct a path.
The basic reconstruction algorithm is as follows:

\begin{algorithm}{Reconstruct Path}

 \multicolumn{2}{|l|}{$\{p_i\}_{i=0,1,2,\cdots}$ = ${ReconstructPath}_{\left(\mathcal{R}(G),s,d,\overline{d}\right)}$~$(g)$} \\
 \multicolumn{2}{|l|}{\!\!\!\!\begin{tabular}{ll} Inputs: &
                                 \!\!\!\!a. The Rips complex of graph, $G=(V,E)$. \\
                               & \!\!\!\!b. Length/cost of the edges, $d: V \rightarrow \mathbb{R}_+$. \\
                               & \!\!\!\!c. The $\overline{d}$-values of the vertices (obtained as output of S* algorithm). \\
                               & \!\!\!\!d. The vertex, $g$, to find the shortest path from.
                              \end{tabular}
                     } \\
 \multicolumn{2}{|l|}{\!\!\!\!\begin{tabular}{ll} Outputs: &
                                 \!\!\!\!a. A sequences of points, $p_i = (\sigma_i, \overline{w}_i)$ for $i=0,1,2,\cdots$.
                              \end{tabular}
                     } \\
 \hline
 
 \newalgline & $\overline{w}_0(g) := 1$ \algcomment{weight map, $\overline{w}: \{g\} \rightarrow (0,1], ~g \mapsto 1$} \\
 \newalgline & $p_0 := (\{g\}, \overline{w}_0)$ \algcomment{$0$-simplex consisting only of $g$, with a weight of $1$ associated with it.} \\
 \newalgline & $\mu := \emptyset$ \\
 \newalgline & $i := 0$ \\
 \newalgline & \textbf{while} $\left(p_i ~\neq~ (\{s\},~ \overline{w}\!:\!\{s\}\!\rightarrow\!(0,1], s\!\mapsto\! 1) ~\right)$ \\
 \newalgline &     \hspace{1.5em} $(p_{i+1}, \mu') := ComputeCameFromPoint_{\left(\mathcal{R}(G),d,\overline{d}\right)} (p_i, \mu)$ \\
  \newalgline &     \hspace{1.5em} $\mu \leftarrow \mu'$ \\
 \newalgline &     \hspace{1.5em} $i \leftarrow i+1$ \\
 \newalgline & \textbf{return} $\{p_i\}_{i=0,1,2,\cdots}$ \\
  
 \hline
\label{alg:attached-maximal-simplices}
\end{algorithm}

\begin{figure}
   \centering
     \subfloat[The point $p_i$ is given by the pair $(\sigma\!=\!\{v_1, v_0, v_3\}, \overline{w})$. $\rho\!=\!\{v_0, v_1, v_2, v_3\}$ is a maximal simplex attached to $\sigma$. It's face, $\gamma \!=\!\{v_0,v_1,v_2\}$, contains the came-from point of $p_i$. To determine the $\overline{d}$-value of $p$, we need to construct the simplex $\gamma \cup \{p\}$, and compute the distance to apex $p$ through that simplex.]{\hspace{0.3in} \includegraphics[width=0.3\columnwidth, trim=5 5 5 5, clip=true]{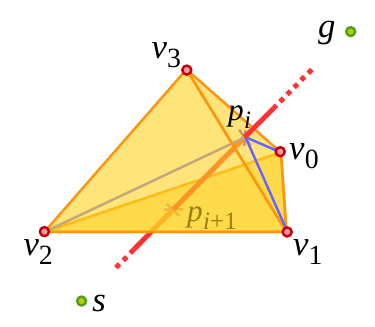} \label{fig:came-from-a} \hspace{0.3in}}
  \caption{Came-from point.}
  \label{fig:came-from}
\end{figure}

The procedure $ComputeCameFromPoint$ takes in a point, $p_i = (\sigma_i, \overline{w}_i)$, on a simplex $\sigma_i$, and returns another point, $p_{i+1} = (\sigma_{i+1}, \overline{w}_{i+1})$ on a different simplex that it \emph{came from} (Figure~\ref{fig:came-from}).
As a result, the relationship between the two simplices, $\sigma_i$ and $\sigma_{i+1}$ is that they both are sub-simplices of a same \emph{maximal simplex}.
Thus, given $p_i$, in order to compute $p_{i+1}$, we construct all the maximal simplices attached to $\sigma_i$ (Line~\ref{line:came-from-pt-maximal-simplex} of Algorithm~\ref{alg:came-from-point}), interpret the point $p_i$ as a vertex in each of the maximal simplices (Lines~\ref{line:p-coord}, \ref{line:d-v-p}), and construct pointed metric simplices using faces of each of them and $p_i$ as apex. The complete pseudocode is as follows:

\begin{algorithm}{Compute Came-From Point in Path Reconstruction}

 \multicolumn{2}{|l|}{$\left( (\sigma',\overline{w}'), \mu' \right)$ = ${ComputeCameFromPoint}_{\left(\mathcal{R}(G),s,d,\overline{d}\right)}$~$\left( (\sigma,\overline{w}), \mu \right)$} \\
 \multicolumn{2}{|l|}{\!\!\!\!\begin{tabular}{ll} Inputs: &
                                 \!\!\!\!a. The Rips complex of graph, $G=(V,E)$. \\
                               & \!\!\!\!b. Length/cost of the edges, $d: V \rightarrow \mathbb{R}_+$. \\
                               & \!\!\!\!c. The $\overline{d}$-values of the vertices (obtained as output of S* algorithm). \\
                               & \!\!\!\!d. A point, $p=(\sigma,\overline{w})$, in the complex. \\
                               & \!\!\!\!e. Maximal simplex, $\mu$, to be ignored.
                              \end{tabular}
                     } \\
 \multicolumn{2}{|l|}{\!\!\!\!\begin{tabular}{ll} Outputs: &
                                 \!\!\!\!a. A point, $p'=(\sigma',\overline{w}')$, in the complex. \\
                               & \!\!\!\!b. The maximal simplex, $\mu'$, through which the line $\overline{p_i p_{i+1}}$ passes.
                              \end{tabular}
                     } \\
 \hline
 
 \newalgline & $d^{ex}:= d$ \algcomment{extended distance map.} \\
 \newalgline & $S := \mathcal{N}(\sigma)$ \algcomment{common neighbors of all the vertices in $\sigma$.} \\
  \newalgline & $d':= \infty$ \algcomment{distance to $p$.} \\
 \newalgline\label{line:came-from-pt-maximal-simplex} & $\mathcal{M}\!S = {MaximalSimplices}_{\mathcal{R}(G)} (S;\sigma)$ \algcomment{maximal simplices attached to $\sigma$} \\
 \newalgline & \textbf{foreach} $(\rho \in \mathcal{M}\!S \!-\! \{\mu\})$ \algcomment{all attached maximal simplices, except $\mu$.} \\
 \newalgline &     \hspace{1.5em} $\overline{w}^{ex} := \overline{w}$ ~; ~~$\overline{w}^{ex}(\nu) := 0, ~\forall \nu\in \rho\!-\!\sigma$ \algcomment{extend domain of $\overline{w}$ to the simplex $\rho$.} \\
 \newalgline &     \hspace{1.5em} $e := \mathcal{E}_d(\rho)$ \algcomment{canonical Euclidean embedding of $\rho$} \\
 \newalgline\label{line:p-coord} &     \hspace{1.5em} $\mathbf{p} := \sum_{v\in\rho} \overline{w}^{ex}(v) ~e(v)$ \algcomment{coordinates of $p=(\sigma,\overline{w})$ in canonical embedding of $\rho$.} \\
 \newalgline\label{line:d-v-p} &     \hspace{1.5em} $d^{ex}(v,p) \!=\! d^{ex}(p,v) := \|e(v) - \mathbf{p}\|, ~\forall v\in \rho$ \algcomment{extend domain of distance/cost function to include $\{v,p\}$.} \\
 \newalgline &     \hspace{1.5em} \textbf{foreach} $(u \in \rho)$ \\
 \newalgline &         \hspace{3em} $\gamma := \rho\!-\!\{u\}$ \algcomment{face of $\rho$ opposite to $u$} \\
 \newalgline &         \hspace{3em} \textbf{if} $(\gamma \nsubseteq \mu)$ \algcomment{ignore any subsimplex of $\mu$.} \\
 \newalgline &             \hspace{4.5em} $\left( d'', (\sigma'',\overline{w}'') \right) := {DistanceThroughSimplex}_{(d^{ex},\overline{d}^{ex})}(\gamma\cup\{p\}, p)$ \\ & \hspace{4.5em}\qquad\qquad\qquad\qquad \algcomment{$\overline{d}$-distance of $p$ through simplex $\gamma\cup\{p\}$ with $p$ as apex.} \\
 \newalgline &             \hspace{4.5em} \textbf{if} $(d'' < d')$ \\
 \newalgline &                 \hspace{6em} $d' \leftarrow d''$ \\
 \newalgline &                 \hspace{6em} $(\sigma',\overline{w}') \leftarrow (\sigma'',\overline{w}'')$ \\
 \newalgline &                 \hspace{6em} $\mu' \leftarrow \rho$ \\
 \newalgline & \textbf{return} $\left( (\sigma',\overline{w}'), \mu' \right)$ \\
 
 \hline
\label{alg:came-from-point}
\end{algorithm}

\section{Analysis}

\subsection{Correctness}


We construct maximal simplices attached to every edge, $\{u,q\}$, in Line~\ref{s-star:attached-maximaal-simplices} of the Basic S* algorithm (Algorithm~\ref{alg:basic-s-star}) and use it for computing and testing for update of the $\overline{d}$-value of a vertex $u$ that is a neighbor to the expanding vertex, $q$.
Since the edge itself is a sub-$1$-simplex of each of those maximal simplices, the distance, $d'$, through the maximal simplex (Line~\ref{s-star:dist-through-simplex}) cannot be greater than the distance obtained by simply adding $d(\{u,q\})$ to $\overline{d}(q)$ (consequence of triangle inequality).
As a result we have the following proposition.

\begin{prop}[$\overline{d}$-value Bounded Above by Graph Search Value]
 The Basic S* algorithm (Algorithm~\ref{alg:basic-s-star}) cannot return a higher value for the distance to any vertex ($\overline{d}$-values) than the Dijkstra's search algorithm.
\end{prop}

Furthermore, a simplicial complex with constant metric inside each simplex gives a piece-wise linear approximation of any Riemannian manifold~\cite{jost:RiemannAnalysis:08}.
The canonical Euclidean embedding of the simplices indeed induces such constant metrics on the simplices (any constant Riemannian metric can be converted to Euclidean metric through appropriate scalings~\cite{petrunin2003polyhedral}). Thus, the proposed method is appropriate for computing shortest paths in discrete representations of Riemannian manifolds.
As a direct consequence, we have the following proposition. A more formal proof is under the scope of future work.

\begin{prop}[$\overline{d}$-value Approaches Path Metric on Riemannian Manifold]
 Suppose $M$ is a Riemannian manifold (possibly with boundaries), and $\mathcal{R}(G)$ is a simplicial approximation 
 of the manifold (with the cost of the edges in $G$ set to their lengths on the manifold). Then the cost/distance between two points on the manifold computed using the Basic S* Algorithm converges to the actual distance between the same points on the manifold as the size of the simplices (lengths/costs of edges in $G$) approach zero.
\end{prop}

\subsection{Complexity}

The Basic S* algorithm (Algorithm~\ref{alg:basic-s-star}) has an overall structure very similar to Dijkstra's algorithm. If the graph, $G$, has $|V|$ counts of vertices, and if all of those are expanded, then the main while block of the algorithm (starting at Line~\ref{s-star:main-loop-start}) will loop for $|V|$ times. Inside each loop, the following processes happen:
\begin{itemize}[itemsep=0em]
 \item[i.] The vertex with lowest $\overline{d}$-value is extracted from set $Q$ (Line~\ref{s-star:extract-q}). The size of $Q$ (open list) is of the order of a constant power of the size of $V$, and since $Q$ is maintained in a heap data structure, the complexity of this step is ~$O(\log |Q|) \sim O(\log |V|^k) \sim O(\log |V|)$.
 \item[ii.] We loop through each neighbor of each vertex to check for updates (Line~\ref{s-star:neighbor-loop}). Assuming average degree of each vertex is $D$, this loops for $O(D)$ times. Inside each of these sub-loops, the following computations happen:
  \begin{itemize}[itemsep=0em]
  \item[a.] The algorithm for computing the set of maximal simplices attached to each edge (Line~\ref{s-star:attached-maximaal-simplices}), as discussed in Section~\ref{sec:attached-maximal}, is of complexity ~$O(|S|^3)\sim O(D^3)$ (where $S$ is the set of neighbors of an edge).
  \item[b.] The duality between maximal simplices and the vertices tells us that the average number of maximal simplices attached to each edge is also $D$. Thus, the innermost loop (starting at Line~\ref{s-star:inner-most-loop} that includes the ``$DistanceThroughSimplex$'' call) loops for $O(D)$ times.
  \end{itemize}
\end{itemize}
Thus, the overall complexity of the algorithm is $O \left(|V| \left( \log(|V|)  +  D^4 + D^2 \right) \right) \sim O \left(|V| \left( \log(|V|)  +  D^4 \right) \right)$.
\subsubsection{Complexity as Size of Configuration Space Increases Keeping Dimension Constant}
If the average degree of the vertices is finite and constant (for a simplicial discretization of a finite-dimensional configuration space that indeed is he case), and if $|V| \rightarrow \infty$, then the complexity is simply $O \left(|V| \log|V| \right)$.
\subsubsection{Complexity as Dimension of Configuration Space Increases Keeping Diameter Constant}
If $N$ is the dimension of the configuration space, and the \emph{diameter} of the configuration space is held constant, then as $N \rightarrow \infty$ we have $|V| \rightarrow O(e^{N}), D \rightarrow O(e^{N})$. Thus, the complexity of the algorithm as $N\rightarrow \infty$ is $O \left(e^{N} \left( \log(e^{N})  +  e^{4N} \right) \right) \sim O (e^{5N})$.


\section{Results}

\begin{figure}
   \centering
     \subfloat[Dijkstra's search: $600$ vertices expanded.]{\fbox{\includegraphics[width=0.22\columnwidth, trim=0 0 0 0, clip=true]{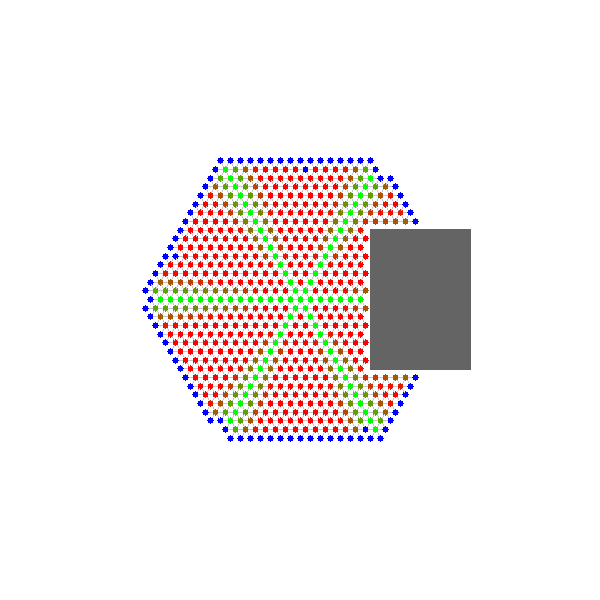}}\label{fig:}}
     \hspace{0.02in}
     \subfloat[$2300$ vertices expanded.]{\fbox{\includegraphics[width=0.22\columnwidth, trim=0 0 0 0, clip=true]{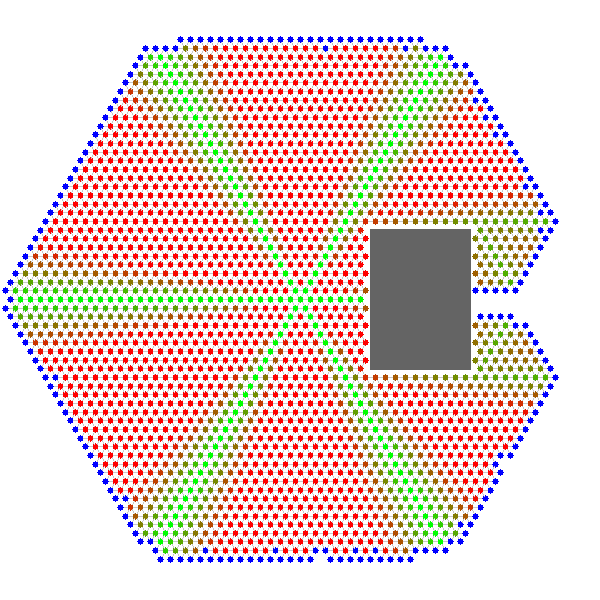}}\label{fig:}}
     \hspace{0.02in}
     \subfloat[all vertices in domain expanded.]{\fbox{\includegraphics[width=0.22\columnwidth, trim=0 0 0 0, clip=true]{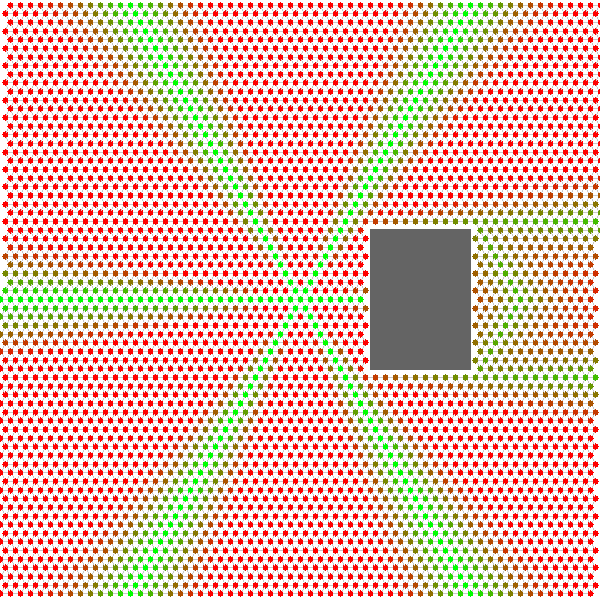}}\label{fig:}}
     \hspace{0.02in}
     \subfloat[Path connecting a vertex to the start.]{\fbox{\includegraphics[width=0.22\columnwidth, trim=0 0 0 0, clip=true]{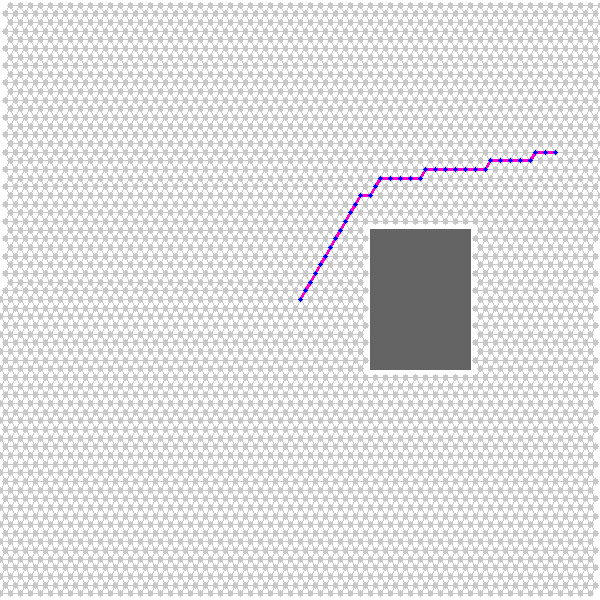}}\label{fig:}}
      \newline
     \subfloat[Basic S* search: $600$ vertices expanded.]{\fbox{\includegraphics[width=0.22\columnwidth, trim=0 0 0 0, clip=true]{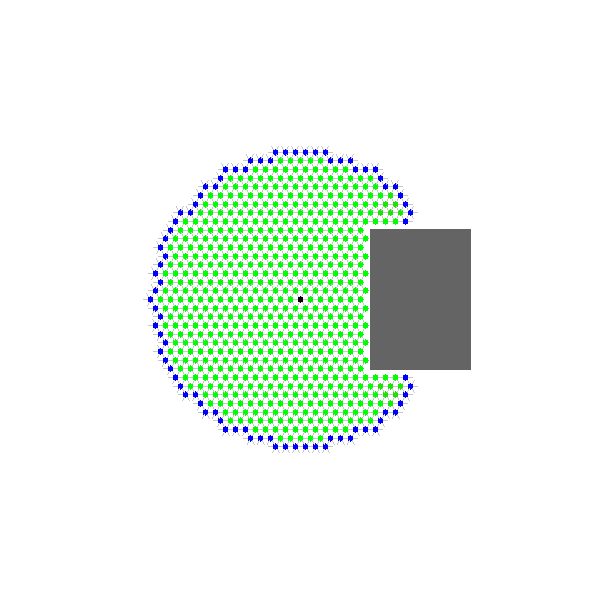}}\label{fig:}}
     \hspace{0.02in}
     \subfloat[$2300$ vertices expanded.]{\fbox{\includegraphics[width=0.22\columnwidth, trim=0 0 0 0, clip=true]{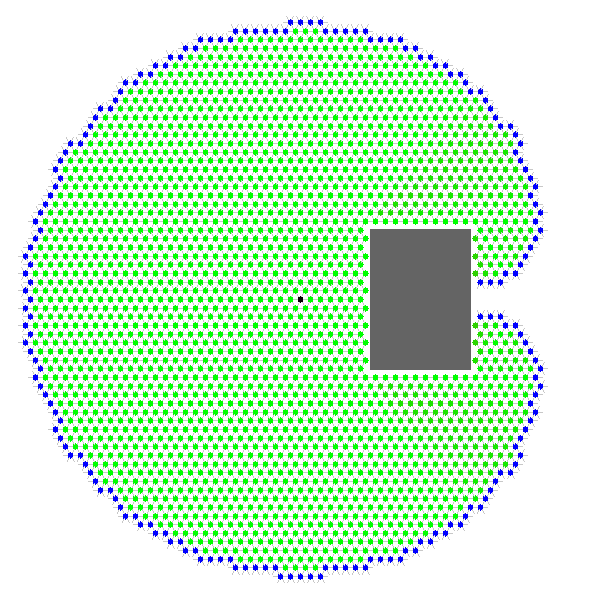}}\label{fig:}}
     \hspace{0.02in}
     \subfloat[all vertices in domain expanded.]{\fbox{\includegraphics[width=0.22\columnwidth, trim=0 0 0 0, clip=true]{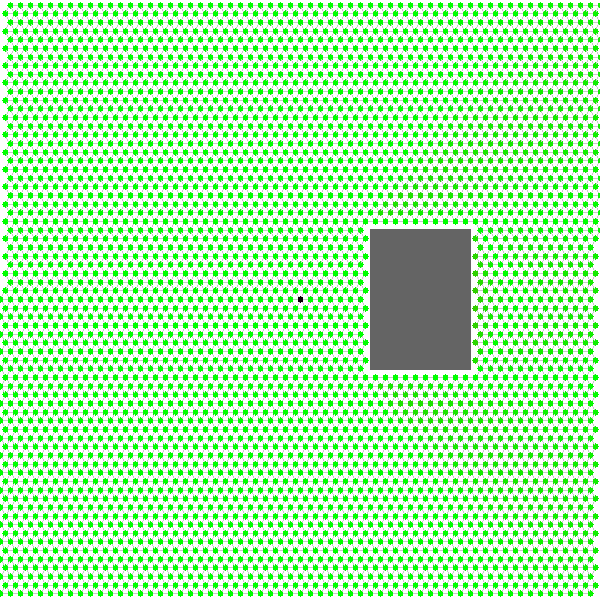}}\label{fig:}}
     \hspace{0.02in}
     \subfloat[Path connecting a vertex to the start.]{\fbox{\includegraphics[width=0.22\columnwidth, trim=0 0 0 0, clip=true]{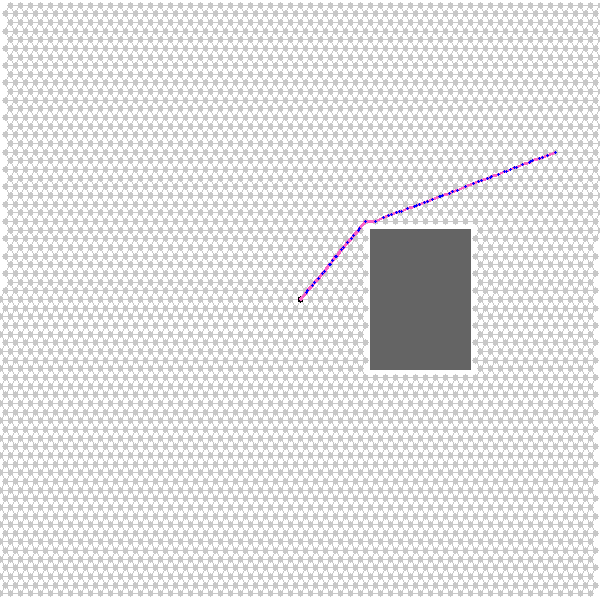}}\label{fig:}}
     \newline
     \subfloat[Vertex color legend for the relative error between the distance computed by search algorithm (figures (a)-(c): Dijkstra's, figures (e)-(g): Basic S*) and the true length of shortest path computed using Euclidean metric.]{\hspace{1.0in} \fbox{\includegraphics[width=0.08\columnwidth, trim=0 0 0 0, clip=true]{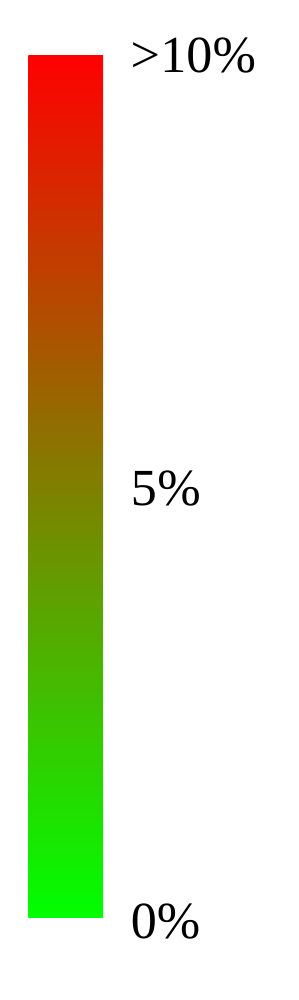}} \hspace{1.0in} \label{fig:}}
     \hspace{0.1in}
     \subfloat[Close-up view of the path computed in (h) using Basic S* search. Blue dots show the points $\{p_i\}_{i=0,1,\cdots}$ returned by the $ReconstructPath$ procedure. Notice how they lie on the boundaries of the maximal $2$-simplices.]{\fbox{\includegraphics[width=0.5\columnwidth, trim=840 840 60 420, clip=true]{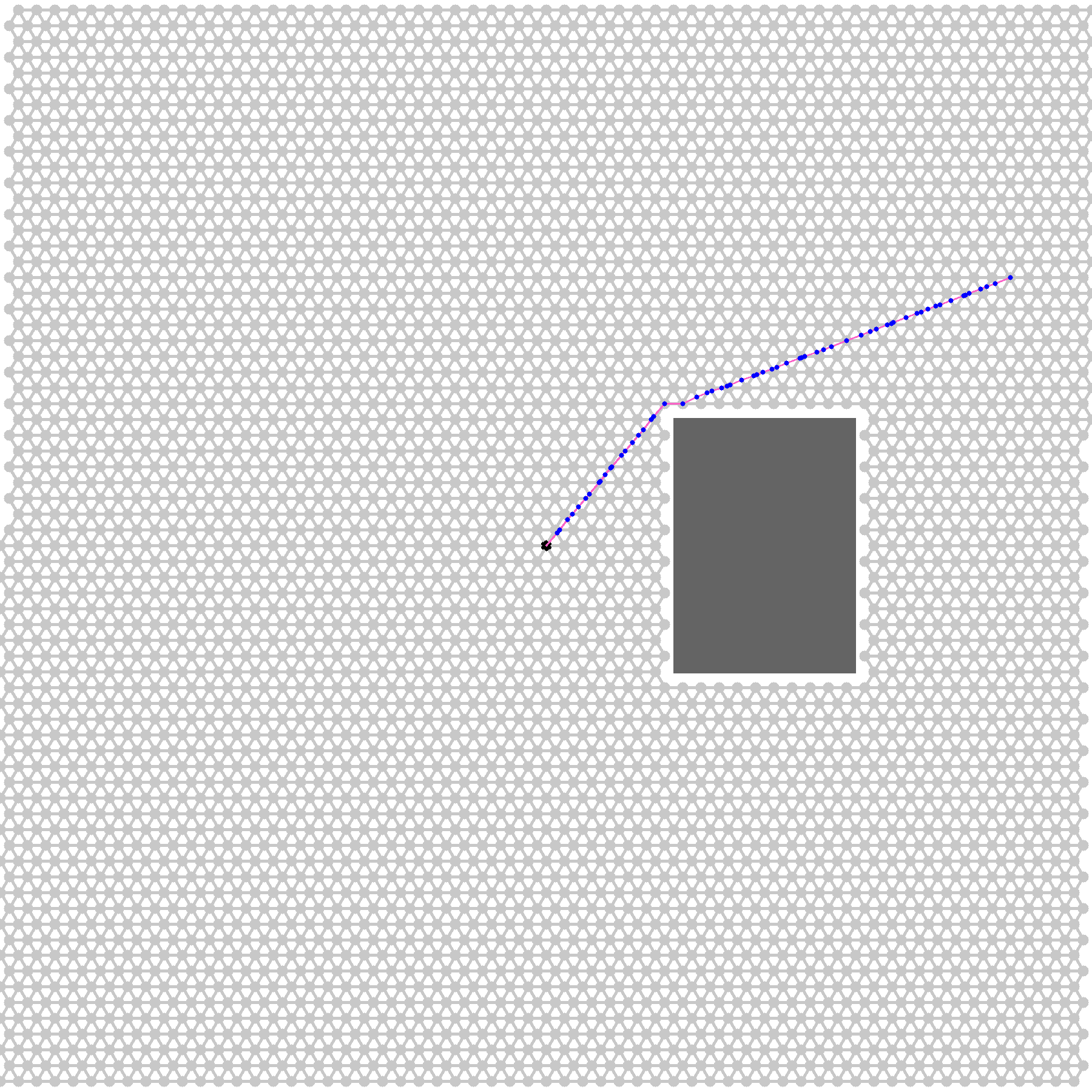}}\label{fig:}}
  \caption{Comparison between Dijkstra's search and Basic S* search using a graph constructed out of an uniform triangulation of an Euclidean plane with a single obstacle. Expanded vertices colored in red and green, while vertices in open list are in blue.
  \emph{Top row:} Progress of Dijkstra's search in (a)-(c). \emph{Second row:} Progress of Basic S* search in (e)-(g).
  Color of the vertices, (i), indicates the relative error between the distance to a vertex as computed by the search algorithm and the true length of shortest path (geodesic distance) to the point using the Euclidean metric on the plane (\emph{green}: low error. \emph{red}: high error). Note how the Basic S* search algorithm computes a distance that more closely represents the underlying path metric induced by the Euclidean metric.}
  \label{fig:rsult-simple1}
\end{figure}

\noindent\textbf{Simple Demonstration:}
As a very simple demonstration, we present a comparison between the progress of search in a graph constructed out of an uniform triangulation (using equilateral triangles) of an Euclidean plane with a single obstacle (Figure~\ref{fig:rsult-simple1}). 

\vspace{0.1in}

\begin{figure}
   \centering
   \begin{tabular}{cc}
     \subfloat[The spherical coordinates.]{{\hspace{0.2in}\includegraphics[width=0.2\columnwidth, trim=0 20 0 0, clip=true]{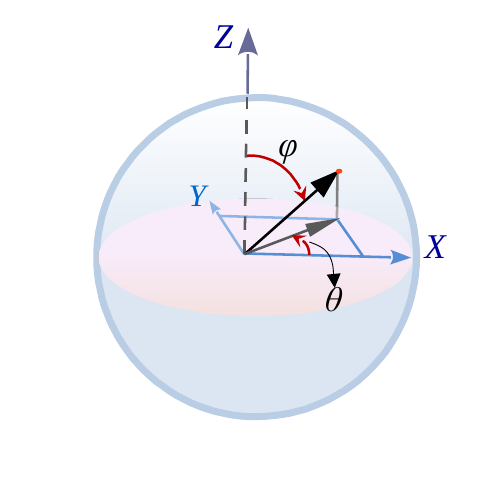}\hspace{0.2in}}\label{fig:result-sphere-coordinates}}
     \hspace{0.05in}
     &
     \multirow{2}{*}[10em]{ 
      \subfloat[The latitude-longitude, $(\phi,\theta)$, coordinate chart. A triangulation graph (light gray) on the coordinate chart, with fineness, $f=8$, is shown. The background color indicates the determinant of the metric tensor, which corresponds to \emph{cost} of edges in a particular region (higher: red, lower: green). 
      ]{\hspace{0.4in} \fbox{\includegraphics[width=0.35\columnwidth, trim=0 0 0 0, clip=true]{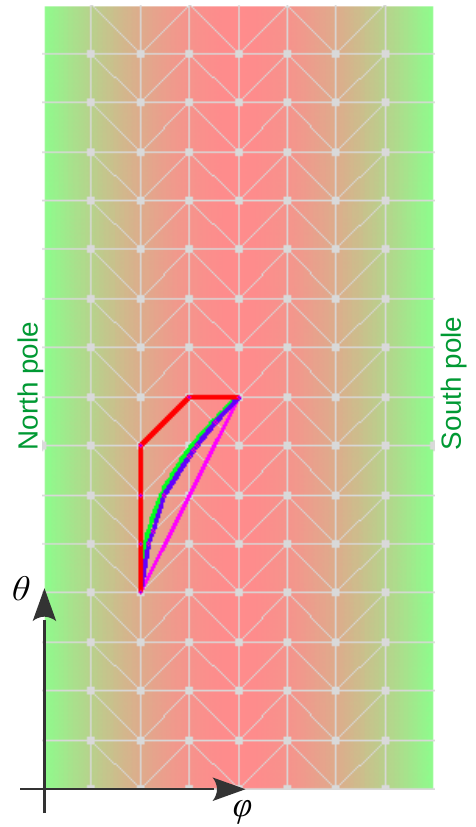}}\label{fig:result-sphere-all-paths} \hspace{0.4in}}
     } \\
     \subfloat[Paths connecting two points on the surface of a sphere. Red paths are computed using Dijkstra's, while blue paths are computed using Basic S*. Multiple paths of each of these colors correspond to different values of fineness, $f$, used in the discretization.]{\hspace{0.3in}\fbox{\includegraphics[width=0.32\columnwidth, trim=0 0 0 0, clip=true]{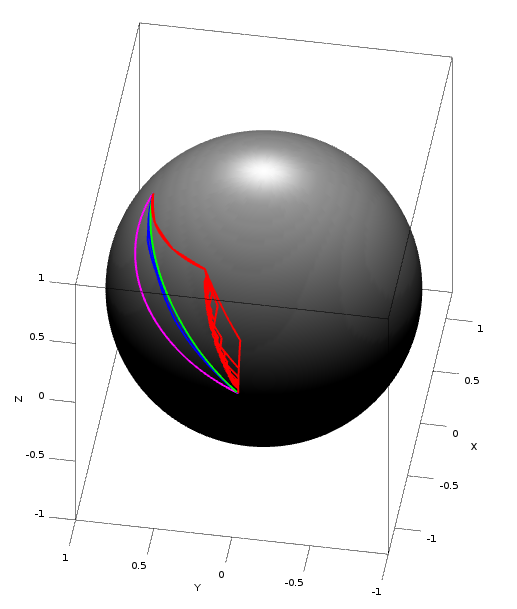}}\label{fig:result-sphere-paths}\hspace{0.3in}} & 
     \\
     \multicolumn{2}{p{\columnwidth}}{\small\subref{fig:result-sphere-paths},\subref{fig:result-sphere-all-paths}: Optimal paths computed on the surface of an unit sphere using Dijkstra's/A* search (red) and Basic S* search (blue). For reference, the figure also shows the geodesic path (green) and the path corresponding to a straight line segment on the $(\phi,\theta)$ coordinate chart (magenta). \subref{fig:result-sphere-all-paths} shows the paths on the $(\phi,\theta)$ coordinate chart computed for a discretization of fineness $f=8$. \subref{fig:result-sphere-paths} shows the paths on the surface of a sphere for various values of fineness, $f=8,16,24,\cdots,80$.}
     \\
     \multicolumn{2}{c}{\subfloat[The total cost/length of paths computed using Dijkstra's (red) and Basic S* (blue) using different fineness values for the discretization (data points at $f=8,16,24,\cdots,80$). The cost of two reference paths (geodesic on sphere: green, straight line segment on coordinate chart: magenta) are also shown.]{\hspace{1in} {\includegraphics[width=0.5\columnwidth, trim=50 200 50 200, clip=true]{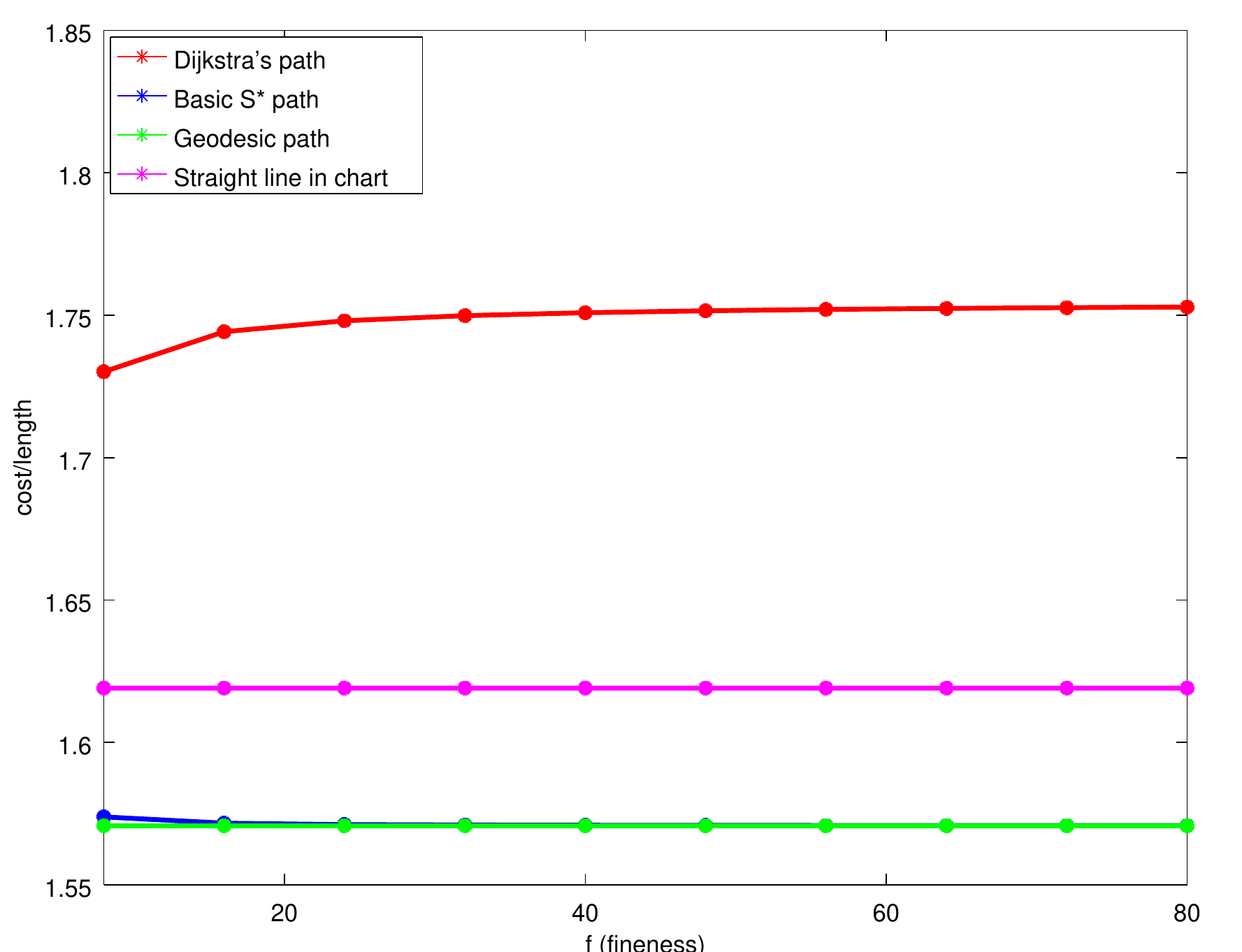}}\label{fig:result-sphere-costs} \hspace{1in}} }
  \end{tabular}
  \caption{Computation of optimal path connecting two points on a $2$-sphere.} \label{fig:result-sphere}
\end{figure}

\noindent\textbf{Shortest Path on a $2$-sphere:}
We next present the result of computing the shortest path (geodesic) on the surface of an unit sphere.
We use the usual spherical coordinates, $(\phi,\theta)$, where $\phi\in[0,\pi]$ is the latitudinal angle measured from the positive $Z$ axis, and $\theta\in[0,2\pi)$ is the longitudinal angle measured from the positive $X$ axis (Figure~\ref{fig:result-sphere}\subref{fig:result-sphere-coordinates}).

The matrix representation of the Riemannian metric tensor~\cite{jost:RiemannAnalysis:08} in this coordinate system is ~$g = \left[ \begin{array}{cc} 1 & 0 \\ 0 & \sin^2 \phi \end{array} \right]$ (the \emph{round metric}). Thus, an infinitesimal segment at $(\phi,\theta)$ of extent $\Delta \phi$ along the $\phi$ direction and $\Delta\theta$ along the $\theta$ direction will be of length/cost $\Delta\l = \sqrt{\Delta\phi^2 + \sin^2\!\phi ~\Delta\theta^2}$.
In particular, using a discrete graph representation (Figure~\ref{fig:result-sphere}\subref{fig:result-sphere-all-paths}) of the coordinate chart, if two neighboring vertices, $v_1$ and $v_2$, have spherical coordinates  $(\phi_1,\theta_1)$ and  $(\phi_2,\theta_2)$ respectively, we compute the cost/length of the edge connecting those vertices as $d({v_1,v_2}) = \sqrt{(\phi_2-\phi_1)^2 + \sin^2\!\left(\frac{\phi_1 + \phi_2}{2}\right) (\theta_2-\theta_1)^2}$.

We construct the graph, $G$, out of a vertex set that has vertices placed on an uniform square lattice, with the separation of the neighboring vertices in the $\phi$ or $\theta$ direction being equal to $\pi/ f$, where $f$ is the ``\emph{fineness}'' of the discretization (larger the value of $f$, finer is the discretization). Figure~\ref{fig:result-sphere}\subref{fig:result-sphere-all-paths} shows such a discretization of the chart with $f=8$.
Comparison of paths computed using Dijkstra's search on this graph and the Basic S* search are shown in Figure~\ref{fig:result-sphere}.
As we increase the fineness value, $f$, it is to be noted that the graphs for lower fineness are not subsets of the graphs of higher fineness. Thus, interestingly, the cost of the paths computed using Dijkstra's search increase with fineness (Figure~\ref{fig:result-sphere}\subref{fig:result-sphere-costs}), but the cost of the paths computed using Basic S* decreases and approaches the geodesic path (great circle) on the sphere.


%

\section{Conclusion}

We presented the \emph{Basic S*} algorithm for computing optimal path through simplical complexes, in particular, Rips complexes of graphs constructed as discrete representation of an arbitrarily configuration space.
The currently proposed algorithm is the basic version of it, with a structure similar to Dijkstra's. However, incorporating heuristic functions, any-time/incremental computations and randomized searches is possible within this framework.
The presented results illustrate the effectiveness of the algorithm in computing paths that more closely represent geodesic paths in some $2$-dimensional configuration spaces. In future, results in $3$-dimensional configuration spaces and comparisons with other ``any-angle'' search algorithms will be provided. Also, more formal proofs of the algorithmic correctness propositions will be given.


\clearpage
\appendix

\section*{\huge{Appendix}}

\section{Euclidean Realization of Metric Simplices}


Given a metric $(n-1)$-simplex, $(\sigma, d)$, one can construct an isometric embedding of the simplex in the Euclidean $(n-1)$-space, $e: \sigma \rightarrow \mathbb{R}^{n-1}$ such that $\|e(v_i) - e(v_j)\| = d_{ij}$ and $\mathbf{v}_i := e(v_i)$ has all its last $n-1-i$ coordinates set to zero. The explicit construction can be described as follows:

Suppose,
\begin{equation*}
\begin{array}{rcccccccl}
 \mathbf{v}_0 =~~ [    & 0,        & 0,              & & \cdots & & & 0  & ] ~~\in \mathbb{R}^{n-1} \\
 \mathbf{v}_1 =~~ [    & \vs_{1,0},  & 0,              & & \cdots & & & 0  & ] \\
 \mathbf{v}_2 =~~ [    & \vs_{2,0},  & \vs_{2,1},        & & \cdots & & & 0  & ] \\
 \vdots~~~~  & & & & & \\
 \mathbf{v}_j =~~ [    & \vs_{j,0},  & \vs_{j,1},  & \cdots,  & \vs_{j,j-1}, & 0, & \cdots & 0  & ] \\
 \vdots~~~~  & & & & & \\
 \mathbf{v}_{n-1} =~~ [    & \vs_{n-1,0},  & \vs_{n-1,1},  & \cdots,  & \vs_{n-1,j-1}, & \vs_{n-1,j}, & \cdots & \vs_{n-1,n-2}  & ] \\
\end{array}
\end{equation*}

From the above, one gets for $0 \leq l<j\leq n-1$,
\begin{equation} \label{eq:dist}
 d_{j,l}^2 ~~=~~ \| \mathbf{v}_j - \mathbf{v}_l \|^2 ~~=~~ \left\{ \begin{array}{ll}
				\sum_{p=0}^{j-1} \vs_{j,p}^2, & \text{when $l=0$} \\
				\sum_{p=0}^{l-1} (\vs_{j,p} - \vs_{l,p})^2 ~+~ \sum_{p=l}^{j-1} \vs_{j,p}^2, & \text{when $0 < l <j\leq n-1$}
				\end{array} \right.
\end{equation}

\subsection{Recursive Formula for Computing $\vs_{j,k}$} \label{ap:L2-embedding}

Using \eqref{eq:dist} we compute $\vs_{j,k}$ for $0 \leq k < j, ~0<j$ as follows:

For $k=0, ~j=1$ clearly,
\begin{equation} \label{eq:j_1}
 \vs_{1,0} = \pm d_{1,0}
\end{equation}

For $k=0,~ 1 < j\leq n-1$,
\begin{equation} \label{eq:k_0}
 \begin{array}{rlcl}
  & d_{j,1}^2 - d_{j,0}^2 & = &  (\vs_{j,0} - \vs_{1,0})^2 ~+~ \sum_{p=1}^{j-1} \vs_{j,p}^2 ~-~ \sum_{p=0}^{j-1} \vs_{j,p}^2 \\
  &  & = & \vs_{1,0}^2 - 2 \vs_{j,0} \vs_{1,0} \\
  \Rightarrow & \vs_{j,0} & = &  \left( d_{j,0}^2 - d_{j,1}^2 + \vs_{1,0}^2 \right) / 2 \vs_{1,0}
 \end{array}
\end{equation}

For $0 < k < j-1,~ 1 < j\leq n-1$,
\begin{equation} \label{eq:k_0_j-1}
\begin{array}{rlcl}
 & d_{j,k+1}^2 - d_{j,0}^2 & = &   \sum_{p=0}^{k} (\vs_{j,p} - \vs_{k+1,p})^2 ~+~ \sum_{p=k+1}^{j-1} \vs_{j,p}^2 ~-~ \sum_{p=0}^{j-1} \vs_{j,p}^2 \\
 &  & = & \sum_{p=0}^{k-1} (\vs_{k+1,p}^2 - 2 \vs_{j,p} \vs_{k+1,p}) ~+~  (\vs_{k+1,k}^2 - 2 \vs_{j,k} \vs_{k+1,k}) \\
 \Rightarrow & \vs_{j,k} & = &  \left( d_{j,0}^2 - d_{j,k+1}^2 + \vs_{k+1,k}^2 + \sum_{p=0}^{k-1} (\vs_{k+1,p}^2 - 2 \vs_{j,p} \vs_{k+1,p}) \right) / 2 \vs_{k+1,k}
 \end{array}
\end{equation}

For $k = j-1,~ 1 < j\leq n-1$,
\begin{equation} \label{eq:k_j-1}
 \begin{array}{rlcl}
 & d_{j,0}^2 & = & \sum_{p=0}^{j-1} \vs_{j,p}^2 \\
 \Rightarrow & \vs_{j,j-1} & = & \pm \sqrt{d_{j,0}^2 - \sum_{p=0}^{j-2} \vs_{j,p}^2}
 \end{array}
\end{equation}

The important property of the formulae in equations~\eqref{eq:j_1}, \eqref{eq:k_0}, \eqref{eq:k_0_j-1} and \eqref{eq:k_j-1} is that the computation of a term in the sequence $\vs_{1,0}, ~\vs_{2,0}, \vs_{2,1}, ~\vs_{3,0}, \vs_{3,1}, \vs_{3,2}, ~\cdots, \vs_{j,0}, \vs_{j,1},\cdots,\vs_{j,j-1}, ~\vs_{j+1,0}, \cdots, \cdots$ requires only the values of the terms appearing before it. Thus, one can compute these values incrementally starting with $\vs_{1,0}$. Furthermore, inserting a new vertex (say the $(n+1)^{th}$ vertex, $v_n$) to a simplex requires us to compute only the new coordinates $\vs_{n,0}, \vs_{n,1}, \cdots, \vs_{n,n-1}$, for the Euclidean realization of the new extended simplex.

With the understanding that $\sum_{p=\alpha}^\beta h(p) = 0$ whenever $\beta < \alpha$, 
equations~\eqref{eq:j_1}--\eqref{eq:k_j-1} can be written more compactly to give the coordinates of the embedded vertices as,
\begin{equation} \label{eq:simplex-eu-embed}
 \vs_{j,k} ~~=~~ \left\{ \begin{array}{ll}
			  { \left( d_{j,0}^2 - d_{j,k+1}^2 + \vs_{k+1,k}^2 + \sum_{p=0}^{k-1} (\vs_{k+1,p}^2 - 2 \vs_{j,p} \vs_{k+1,p}) \right) } ~/~ { 2 \vs_{k+1,k} }, & \text{for $k < j - 1$}, \\
                          \pm \sqrt{d_{j,0}^2 - \sum_{p=0}^{j-2} \vs_{j,p}^2}, & \text{for $k = j - 1$.} \\
                          0, & \text{for $k \geq j$.} \\
                         \end{array} \right.
\end{equation}

We choose the positive solution for coordinates $\vs_{j,j-1}$.
The computation of $\vs_{1,0}, ~\vs_{2,0}, \vs_{2,1}, ~\vs_{3,0}, \vs_{3,1}, \vs_{3,2}, ~\cdots,$ $\vs_{j,0}, \vs_{j,1},\cdots,\vs_{j,j-1}, ~\vs_{j+1,0}, \cdots, \cdots$ can be made in an incremental manner, with the computation of a term in this sequence requiring only the previous terms.

\begin{lemma}
 Equation~\eqref{eq:simplex-eu-embed} does not have a real solution iff the metric simplex, $(\sigma,d)$, is not Euclidean realizable.
\end{lemma}

\begin{lemma}
 If $\vs_{j,j-1} = 0$ for some $j$, and all the coordinates are real, then $(\sigma,d)$ has a degenerate Euclidean realization.
\end{lemma}


\subsection[Coordinate Computation for a Point with given Distances to non-apex Vertices]{Coordinate Computation for a Point with given Distances to $\{\mathbf{v}_j\}_{j=1,2\cdots,n\!-\!1}$} \label{ap:L2-embedding-o}

Given a metric $(n-1)$-simplex, $(\sigma,d)$, and the canonical Euclidean realization $\mathcal{E}_d(\sigma)=e:v_i \mapsto \mathbf{v}_i$,
we consider an additional point, $o$, and its Euclidean realization in the same Euclidean space, 
\begin{equation}
 \mathbf{o} ~=~~   [ ~\os_{0},  ~\os_{1},  \cdots\cdots, ~\os_{n-2}~ ] ~~\in \mathbb{R}^{n-1}
\end{equation}
%
%
Along with given its distances to all vertices in $\sigma$, except $v_0$: $\overline{d}: \sigma - v_0 \rightarrow \mathbb{R}_{+} $. For convenience, we write $\overline{d}(v_i) = \overline{d}_i, ~i=1,2,\cdots,n-1$. Thus, for $0<l \leq n-1$
\begin{equation}
 \overline{d}_l^2 ~~=~~ \sum_{p=0}^{l-1} (\os_p - \vs_{l,p})^2 ~+~ \sum_{p=l}^{n-2} \os_p^2
\end{equation}

Thus, for $l=2,3,\cdots,n-1$,
\begin{equation}\begin{array}{rrcl}
 & \overline{d}_l^2 - \overline{d}_1^2 & = & \sum_{p=0}^{l-1} (\os_p - \vs_{l,p})^2 + \sum_{p=l}^{n-2} \os_p^2 ~-~ (\os_0 - \vs_{1,0})^2 - \sum_{p=1}^{n-2} \os_p^2 \\
 & & = & -2 \os_0 \vs_{l,0} + \vs_{l,0}^2 + 2 \os_0 \vs_{1,0} - \vs_{1,0}^2 + \sum_{p=1}^{l-1} (-2 \os_p \vs_{l,p} + \vs_{l,p}^2)\\
 & & = & 2(\vs_{1,0} - \vs_{l,0}) \os_0 ~-~ 2 \sum_{p=1}^{l-1} \vs_{l,p} \os_p ~+~ \left( \sum_{p=0}^{l-1} \vs_{l,p}^2 ~-~ \vs_{1,0}^2 \right) \\
\Rightarrow &  \os_{l-1} & = & \frac{\vs_{1,0} - \vs_{l,0}}{\vs_{l,l-1}} \os_0 ~-~  \sum_{p=1}^{l-2} \frac{\vs_{l,p}}{\vs_{l,l-1}} \os_p ~+~ \frac{1}{2 \vs_{l,l-1}}\left( \sum_{p=0}^{l-1} \vs_{l,p}^2 ~-~ \vs_{1,0}^2 +  \overline{d}_1^2 - \overline{d}_l^2 \right) \\
\Rightarrow &  \os_{k} & = & \alpha_{k,0} \os_0 ~+~ \sum_{p=1}^{k-1} \alpha_{k,p} \os_p ~+~ \beta_k \qquad \text{(with $k=l-1$)}
\end{array} \end{equation}
where, $\alpha_{k,0} = \frac{\vs_{1,0} - \vs_{k+1,0}}{\vs_{k+1,k}}$, 
~~$\alpha_{k,p} = - \frac{\vs_{k+1,p}}{\vs_{k+1,k}}, ~p=1,2,\cdots,k-1$ 
and ~~$\beta_k =  \frac{1}{2 \vs_{k+1,k}}\left( \sum_{p=0}^{k} \vs_{k+1,p}^2 ~-~ \vs_{1,0}^2 +  \overline{d}_1^2 - \overline{d}_{k+1}^2 \right)$.


Letting
\begin{equation}
 \os_l = A_l \os_0 + B_l 
\end{equation}
we have,
\begin{equation}\begin{array}{rcl}
 \os_k & = & \alpha_{k,0} \os_0 ~+~ \sum_{p=1}^{k-1} \alpha_{k,p} \os_p ~+~ \beta_k \\ 
       & = & \alpha_{k,0} \os_0 ~+~ \sum_{p=1}^{k-1} \alpha_{k,p} (A_p \os_0 + B_p) ~+~ \beta_k \\
       & = & \left( \alpha_{k,0} + \sum_{p=1}^{k-1} \alpha_{k,p} A_p  \right) \os_0 ~+~ \left( \beta_k + \sum_{p=1}^{k-1} \alpha_{k,p} B_p \right)
\end{array}\end{equation}

Thus, $A_k$ and $B_k$ can be determined iteratively as follows:
\begin{equation}\begin{array}{ll}
 A_1 = \alpha_{1,0},~~~ & B_1 = \beta_1 \\
 A_k = \alpha_{k,0} + \sum_{p=1}^{k-1} \alpha_{k,p} A_p,~~~ & B_k = \beta_k + \sum_{p=1}^{k-1} \alpha_{k,p} B_p
\end{array}\end{equation}

We can now determine $\os_0$ from the expression for $\overline{d}_1^2$:
\begin{equation}\begin{array}{rl}
  & \begin{array}{rcl}
  \overline{d}_1^2 & = & (\os_0 - \vs_{1,0})^2 + \sum_{p=1}^{n-2} \os_p^2 \\
                  & = & ( \os_0^2 - 2 \vs_{1,0}\os_0 +  \vs_{1,0}^2) ~+~ \sum_{p=1}^{n-2} (A_p^2 \os_0^2 + 2 A_p B_p \os_0 + B_p^2)
  \end{array} \\
\Rightarrow & \left( 1 + \sum_{p=1}^{n-2} A_p^2 \right) \os_0^2 ~+~ 2 \left( -\vs_{1,0} + \sum_{p=1}^{n-2} A_p B_p \right) \os_0 ~+~ \left( \vs_{1,0}^2 - \overline{d}_1^2 + \sum_{p=1}^{n-2} B_p^2 \right) ~~=~~ 0
\end{array}\end{equation}


We need to choose the solution of $\mathbf{o}$ such that $\mathbf{v}_0$ and $\mathbf{o}$ lies on the opposite sides of the hyperplane containing the points $\mathbf{v}_1, \mathbf{v}_2, \cdots, \mathbf{v}_{n-1}$.
Since we have chosen $\vs_{1,0}$ to be positive, and since $\mathbf{v}_0$ is the origin, choosing the higher value of $\os_0$ will satisfy this condition. Thus,
\begin{equation} \label{eq:o0}
 \os_0 = \frac{-V+\sqrt{V^2 - 4UW}}{2U}
\end{equation}
where, $U = \left( 1 + \sum_{p=1}^{n-2} A_p^2 \right), ~V = 2 \left( -\vs_{1,0} + \sum_{p=1}^{n-2} A_p B_p \right), ~W = \left( \vs_{1,0}^2 - \overline{d}_1^2 + \sum_{p=1}^{n-2} B_p^2 \right)$

\begin{lemma}
 \eqref{eq:o0} does not have a real solution iff $o$ is not embeddable in the same Euclidean space as an Euclidean realization of $(\sigma,d)$ satisfying the distance relations $\overline{d}$.
\end{lemma}


\subsection{Intersection Point of Plane and Line Segment} \label{ap:L2-embedding-line-plane-intersection}

A general point on the (affine) hyperplane, $H_0$, containing points $\mathbf{v}_1, \mathbf{v}_2, \cdots, \mathbf{v}_{n-1}$ can be written as $\sum_{i=1}^{n-1} w_i \mathbf{v}_i$, where $\sum_{i=1}^{n-1} w_i = 1$. A point on the line, $L$, joining $\mathbf{v}_0$ and $\mathbf{o}$ can be written as $w_0 \mathbf{v}_0 + w_o \mathbf{o} ~( = w_o \mathbf{o})$ (since $\mathbf{v}_0 = [0,0,\cdots,0]$), where $w_0 + w_o = 1$.
Thus, the point of intersection of $H_0$ and $L$ is
\begin{eqnarray} \label{eq:weights}
& & \mathbf{i}_0 ~=~ \sum_{i=1}^{n-1} w_i \mathbf{v}_i ~=~ w_o \mathbf{o} \nonumber \\
\Rightarrow & & \sum_{i=1}^{n-1} w_i \vs_{i,j-1} ~=~ w_o \os_{j-1}, ~~~~\text{for } j=1,2,\cdots,n-1 \nonumber \\
\Rightarrow & & \sum_{i=j}^{n-1} w_i \vs_{i,j-1} ~=~ w_o \os_{j-1}  ~~~~\text{(since $\vs_{i,j-1}=0$ for $i\leq j-1$)} \nonumber \\
\Rightarrow & & w_j = \frac{ w_o \os_{j-1} - \sum_{i=j+1}^{n-1} w_i \vs_{i,j-1} }{\vs_{j,j-1}} \nonumber \\
\Rightarrow & & {w'}_j = \frac{ \os_{j-1} - \sum_{i=j+1}^{n-1} {w'}_i \vs_{i,j-1} }{\vs_{j,j-1}}, ~~~~\text{where, } {w'}_j = \frac{w_j}{w_o}
\end{eqnarray}
The above gives a recursive formula that lets us compute the terms in the sequence ${w'}_{n-1}, {w'}_{n-1}, \cdots, {w'}_1$ in an incremental manner, with computation of each term requiring the knowledge of the previous terms in the sequence only.

Since $\sum_{i=1}^{n-1} w_i = 1$ and ${w'}_j = \frac{w_j}{w_o}$, we have the following
\begin{equation}
 w_j = \frac{{w'}_j}{\sum_{i=1}^{n-1} {w'}_i}
\end{equation}


\subsubsection{Sign of the Wights}

In general, let $H_{k_1,k_2,\cdots,k_p}$ be the $(n-1-p)$-dimensional hyperplane containing all the points in the set $\{\mathbf{v}_i\}_{i=0,1,\cdots,n-1} - \{\mathbf{v}_{k_j}\}_{j=1,2,\cdots,p}$ (\emph{i.e.,} the hyperplane of the subsimplex not containing $\mathbf{v}_{k_1}, \mathbf{v}_{k_2}, \cdots, \mathbf{v}_{k_p}$).

\begin{lemma}
\begin{equation}
 w_k ~~\left\{ \begin{array}{l} 
                         < 0 ~~~\text{iff $\mathbf{i}_0$ and $\mathbf{v}_k$ lie on the opposite sides of $H_{0,k}$ in $H_0$} \\
                         = 0 ~~~\text{iff $\mathbf{i}_0$ lies on $H_{0,k}$} \\
                         > 0 ~~~\text{iff $\mathbf{i}_0$ and $\mathbf{v}_k$ lie on the same sides of $H_{0,k}$ in $H_0$} \\
                        \end{array} \right.
\end{equation}
\end{lemma}

\subsection[Computation of Plane with Given Distances from non-apex Vertices]{Computation of Plane with Given Distances from $\{\mathbf{v}_j\}_{j=1,2\cdots,n\!-\!1}$} \label{ap:dist-from-plane}

Given a metric $(n-1)$-simplex, $(\sigma,d)$, and the canonical Euclidean realization $\mathcal{E}_d(\sigma)=e:v_i \mapsto \mathbf{v}_i$,
we consider a hyperplane, $I$, described by the equation, $\mathbf{u}\cdot \mathbf{x} + \mu = 0$, where, $\mathbf{x}\in\mathbb{R}^{n-1}$ is a point on the hyperplane, $\mathbf{u} = [\us_{0}, \us_{1}, \cdots, \us_{n-2}] \in\mathbb{R}^{n-1}$ is an unit vector orthogonal to the plane, and $\mu$ is a constant.

Distance of the point $\mathbf{v}_j$ from the plane is $\overline{d}_j$. Thus, for $j=1,2,\cdots,n\!-\!1$, we have,
\begin{eqnarray} \label{eq:plane-dists}
 & & \mathbf{u}\cdot \mathbf{v}_j + \mu = \overline{d}_j  \\
 \Rightarrow & & \sum_{p=0}^{j-1} \us_p \vs_{j,p} ~+~ \mu ~=~ \overline{d}_j \nonumber \\
 \Rightarrow & & \us_{j-1} ~=~ \frac{\overline{d}_j - \mu - \sum_{p=0}^{j-2} \us_p \vs_{j,p}}{\vs_{j,j-1}} \nonumber \\
 \Rightarrow & & \us_{k} ~=~ \frac{\overline{d}_{k+1} - \mu - \sum_{p=0}^{k-1} \us_p \vs_{k+1,p}}{\vs_{k+1,k}} \nonumber
\end{eqnarray}
Letting
\begin{equation} \label{eq:plane-u-M-N}
 \us_{l} ~=~ M_{l}~\mu + N_{l}
\end{equation}
we have
\begin{eqnarray}
 \us_{k} & = & \frac{\overline{d}_{k+1} - \mu - \sum_{p=0}^{k-1} (M_p\mu + N_p) \vs_{k+1,p}}{\vs_{k+1,k}} \nonumber \\
         & = & -\left( 1 + \frac{1}{\vs_{k+1,k}} \sum_{p=0}^{k-1} M_p \vs_{k+1,p} \right) \mu ~+~ \left( \overline{d}_{k+1} - \frac{1}{\vs_{k+1,k}}\sum_{p=0}^{k-1} N_p \vs_{k+1,p}  \right) 
\end{eqnarray}
Thus, we have the following recursive equation for $M_k$ and $N_k$,
\begin{equation} \label{eq:plane-M-N}
 M_k = -\left( 1 + \frac{1}{\vs_{k+1,k}} \sum_{p=0}^{k-1} M_p \vs_{k+1,p} \right), \qquad
 N_k = \left( \overline{d}_{k+1} - \frac{1}{\vs_{k+1,k}}\sum_{p=0}^{k-1} N_p \vs_{k+1,p}  \right)
\end{equation}
for $k=0,1,2,\cdots,n-2$. With the understanding that $\sum_{p=\alpha}^\beta h(p) = 0$ whenever $\beta < \alpha$, we have $M_0 = -1, ~N_0 = \overline{d}_1$.

Since $\mathbf{u}$ is an unit vector, $\sum_{j=0}^{n-1} \us_j^2 = 1$. Thus,
$\left( \sum_{j=0}^{n-1} M_j^2 \right) \mu^2 + \left( 2 \sum_{j=0}^{n-1} M_j N_j \right) \mu + \left( \sum_{j=0}^{n-1} N_j^2 - 1 \right) = 0$.
Thus,
\begin{equation} \label{eq:plane-mu}
  \mu = \frac{-Q + \sqrt{Q^2 - 4PR}}{2P}
\end{equation}
where, $P = \sum_{j=0}^{n-1} M_j^2, ~Q = 2 \sum_{j=0}^{n-1} M_j N_j$ and $R = \sum_{j=0}^{n-1} N_j^2 - 1$. We choose the positive sign before the square root since we want the plane satisfying~\eqref{eq:plane-dists} that is farthest from the origin.

Equations \eqref{eq:plane-M-N}, \eqref{eq:plane-mu} and \eqref{eq:plane-u-M-N} computes the required plane, $I$, which is at a distance $\overline{d}_j$ from $\mathbf{v}_j$, $j=1,2,\cdots,n\!-\!1$.

Similar to what described in Section~\ref{ap:L2-embedding-line-plane-intersection}, the point at which the perpendicular (in the direction of $\mathbf{u}$) dropped from $\mathbf{v}_0$ onto the hyperplane $I$ intersects the hyperplane $H_0$ is given by
$\mathbf{i}_0 = \sum_{i=1}^{n-1} w_i \mathbf{v}_i$, with
\begin{equation} \label{eq:weights-spherical}
 w_k = \frac{{w'}_k}{\sum_{i=1}^{n-1} {w'}_i}, ~~~~k=1,2,\cdots, n\!-\!1.
\end{equation}
where, ${w'}_j$ can be computed recursively using the formula ${w'}_j = \frac{\us_{j-1} - \sum_{i=j+1}^{n-1} {w'}_i \vs_{i,j-1} }{\vs_{j,j-1}}$.

%



\bibliographystyle{alpha}
\bibliography{s_star}
\end{document}